\documentclass[aps,pra,showpacs,twoside,10pt,floatfix,nofootinbib,longbibliography,twocolumn]{revtex4-1}
\usepackage[colorlinks=true, citecolor=blue, urlcolor=blue]{hyperref}
\usepackage{epsfig,newlfont,amssymb,amsfonts,amsmath,bm,subfigure,palatino,mathtools,amsthm,braket,soul,enumitem,color,graphics,graphicx,times,physics,bbold,comment, braket, ulem, qcircuit}
\usepackage[thinc]{esdiff}
\usepackage{xcolor}
\usepackage{dsfont}
\usepackage{float}
\usepackage{tabularx}
\usepackage{array}
\usepackage[export]{adjustbox}
\usepackage{mathtools}
\newtheorem{theorem}{Theorem}
\newtheorem{definition}{Definition}

\newtheorem{lemma}{Lemma}
\newtheorem{corollary}[theorem]{Corollary}
\newtheorem{proposition}[theorem]{Proposition}

\newcommand{\ga}[1]{{\color{blue}#1}}

\newcolumntype{Y}{>{\centering\arraybackslash}X}
\begin{document}

\title{Quantum time-flip beats adaptive metrology: Asymptotic benefit, activation, and unsimulability}

\author{Gaurang Agrawal$^{1,2,3, 5}$, Pritam Halder$^{2,3,4}$, Aditi Sen(De)$^{2,3}$}
\affiliation{$^1$Indian Institute of Science Education and Research, Homi Bhabha Rd, Pashan, Pune 411 008, India\\
$^2$Harish-Chandra Research Institute, Chhatnag Road, Jhunsi, Prayagraj - 211019, India\\
$^3$Homi Bhabha National Institute, Training School Complex, Anushakti Nagar, Mumbai 400094, India\\
\(^4\)Networked Quantum Devices Unit, Okinawa Institute of Science and Technology Graduate University, Okinawa, Japan\\
$^5$ Université Paris-Saclay, Inria, CNRS, LMF, 91190 Gif-sur-Yvette, France}

\begin{abstract}

In quantum metrology, adaptive and causal-superposition strategies  are {proven} to be beneficial over parallel schemes for a finite number of channel uses, but their advantages  disappear in the asymptotic limit. We show that quantum operations with indefinite time direction, specifically, time-flip (TF)-assisted strategies,  referred to as indefinite time-directed metrology (ITDM), can overcome this asymptotic equivalence. Using semidefinite programming, we rigorously demonstrate that TF-assisted protocols can achieve quantum Fisher information (QFI) strictly exceeding the maximum value attainable by parallel, adaptive, and causal-superposition strategies, both for finite and asymptotically many channel uses. Moreover, we identify a class of Pauli noise channels for which ITDM achieves Heisenberg scaling, while all parallel, adaptive, and causal-superposition strategies remain restricted to standard scaling. We call  this phenomenon as metrological activation. Interestingly, this activation can be used to  exhibit that the quantum time-flip and transposition supermaps cannot be simulated by conventional quantum circuits or causal-superposition strategies using any finite number of channel queries, thereby establishing indefinite time direction as a genuine resource for quantum metrology.

\end{abstract}

\maketitle

\section{Introduction}

\begin{figure} 
    \centering
\includegraphics[width=0.5\textwidth]{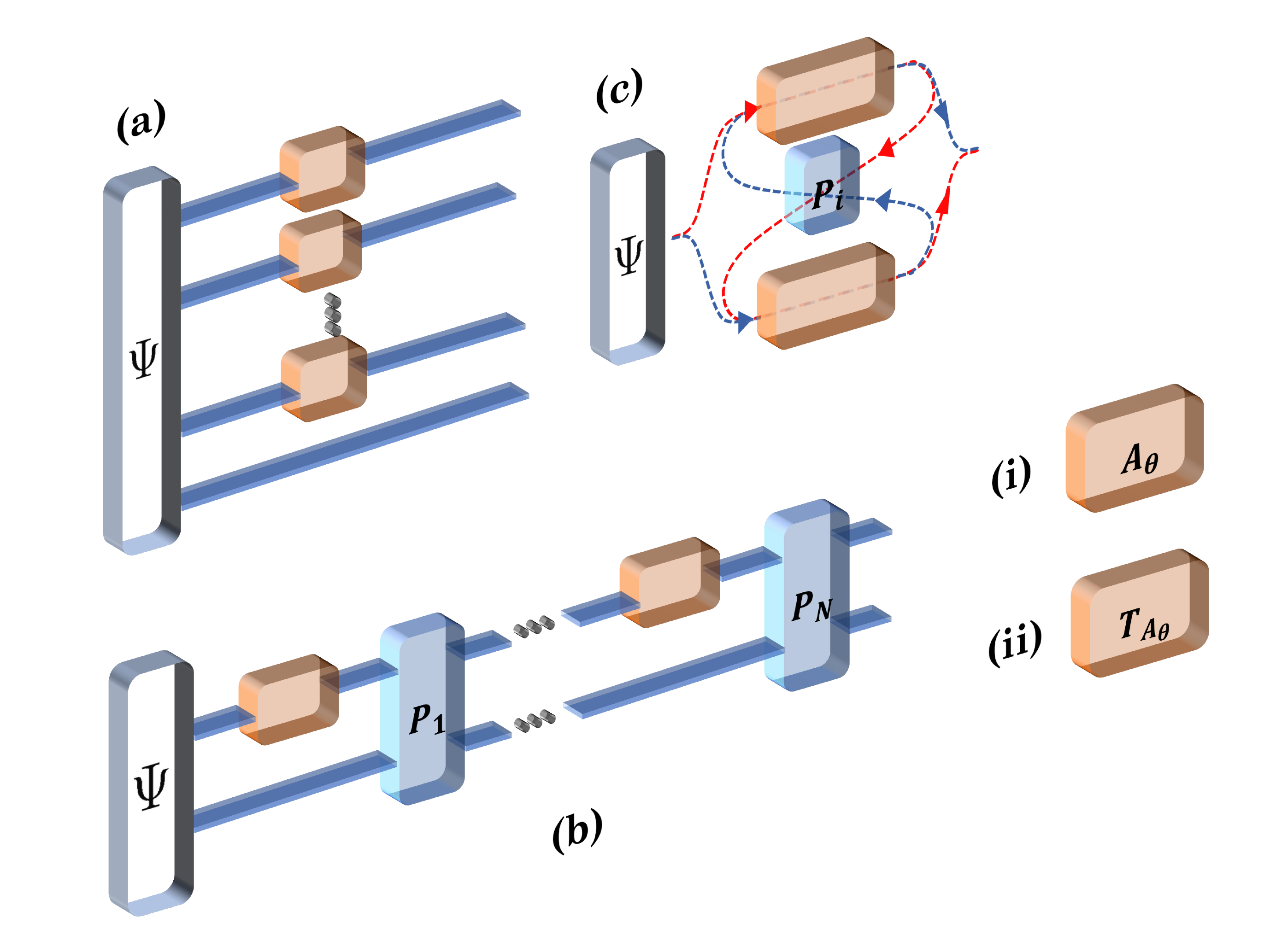}
    \caption{\textbf{ Schematic of various strategies.} (a) A parallel strategy ({\bf PAR}) is presented. $N$ parametrized channels act on the input probe $\ket{\Psi}$ to create an encoded probe $\rho_\theta$. (b) In an adaptive strategy ({\bf AD}), a feedback or  correction isometry $P_i$ is applied after every round of the channel. (c)  Orders of $P_i$ are controlled by a quantum degree of freedom in the  causal superposition strategy ({\bf CS}).  In all these scenarios, either regular (i) $ A_\theta^{\otimes N}$ or TF-assisted (ii) $T_{A_\theta}^{\otimes N}$ metrology can be performed.} 
    \label{fig:Par_AD_CS}
\end{figure}

The central goal of quantum metrology~\cite{Giovannetti2004Nov, Giovannetti2011Apr} is to utilize the principles of  quantum mechanics to enhance the precision of estimating physical parameters. Remarkable demonstrations of this quantum advantage have been achieved in applications ranging from quantum imaging~\cite{Perez-Delgado2012Sep, Genovese2016Jun, Moreau2019Jun} and atomic clocks~\cite{Appel2009Jul,Katori2011Apr,Ludlow2015Jun,Nichol2022Sep} to magnetometry~\cite{Jones2009Apr,Wasilewski2010Mar}, gravitational-wave detection~\cite{Caves1981Apr,Schnabel2010Nov,BibEntry2011Dec}, and quantum radar~\cite{Barzanjeh_2015}. Given $N$ queries of the parametrized quantum channel, a central challenge is to surpass the standard scaling (SS), with precision proportional to $1/\sqrt{N}$, and attain the Heisenberg scaling (HS), proportional to $1/N$. It was shown that the quantumness of a probe, like quantum coherence, can enable HS under repeated channel applications in a noiseless scenario~\cite{Higgins2007,Braun_2018}. In realistic settings, however, unavoidable system-environment interactions introduce noise that can suppress quantum advantage, leading to SS. This has motivated a hierarchy of metrological strategies (see Fig.~\ref{fig:Par_AD_CS}),  ranging from parallel protocols (\textbf{Par}), in which probes and channels are employed simultaneously~\cite{Giovannetti2006Jan}, to adaptive protocols (\textbf{AD}), where the channels are used sequentially with active feedback in between via error correction or quantum control on the probe states~\cite{vanDam2007Mar,Demkowicz-Dobrzanski2014Dec,Demkowicz-Dobrzanski2017Oct,Sekatski2017Sep,Pirandola2017Mar}. The most general processes with a definite causal structure, including adaptive protocols, can be described within the framework of quantum combs~\cite{Chiribella2008Aug,Chiribella2009Aug}. Going beyond definite causal order, indefinite causal order ({\bf ICO}) protocols, and their subclasses like causal superpositions (\textbf{CS}) such as the quantum switch, coherently superpose different causal sequences~\cite{chiribella2013quantum,Oreshkov2012Oct,Liu2024Dec}. Such approaches have been shown to offer metrological advantages, with quantum switch-assisted protocols exhibiting super-Heisenberg scaling in continuous-variable settings~ \cite{zhao2020quantum, Yin2023Aug}. More recently, indefinite time-directed quantum operations have introduced a distinct resource: the quantum time-flip (\textbf{TF}) supermap~\cite{chiribella2022quantum}, which coherently probes a channel in forward and backward time directions by interchanging its input and output. With the aid of quantum time-flip supermap, we
have demonstrated that HS can be achieved~\cite{Agrawal2025Jul} without entanglement, along with improved Fisher information in certain noisy scenarios.ntanglement. However, these advantages have so far been established against conventional metrological strategies and not necessarily against optimal ones, raising the fundamental question of whether indefinite time direction can provide a genuine advantage against optimal \textbf{AD} and \textbf{PAR} strategies.

Quantum systems inevitably interact with their environment, thereby reducing the achievable metrological precision. In the presence of such noise, and for a finite number of channel queries, \textbf{ICO} protocols can provide higher precision than the \textbf{AD} strategies, which, in turn, can surpass \textbf{PAR} protocols in certain settings~\cite{liu2023optimal, mothe2024reassessing, Liu2024Dec, Kurdzialek2023Aug}. However, a no-go result, known as the asymptotic equivalence, establishes that in the limit $N\rightarrow\infty$, neither {\bf AD} nor {\bf CS} strategies can offer an advantage over a simple {\bf PAR} one~\cite{Kurdzialek2023Aug}. Thus, the gain of such strategies is ultimately washed out in the asymptotic regime. 

In this work, we ask whether indefinite time direction can overcome this fundamental limitation and answer this query affirmatively.  To address this,  we study  $N$ uses of time-flipped parametrized channels, assisting the various known strategies, including  \textbf{PAR}, \textbf{AD}, and \textbf{CS} which we refer to as \textbf{TF}-assisted metrology. In the finite usage regime, we adopt semidefinite-programming techniques~\cite{liu2023optimal, mothe2024reassessing, Liu2024Dec, Kurdzialek2025Jan, Dulian2025Jun} to demonstrate  the superiority  of \textbf{TF}-assistance over unassisted definite-causal and causal-superposition strategies, as well as over \textbf{ICO} protocols. More importantly, in the asymptotic regime, we show that although the TF-assisted \textbf{PAR}, \textbf{AD}, and \textbf{CS} become equivalent to one another, they retain a strict advantage over their unassisted counterparts. Thus, TF-assistance overcomes the asymptotic no-go theorem and maintains a metrological advantage even as $N\rightarrow\infty$. We further identify estimation tasks for which Heisenberg scaling cannot be achieved by employing   any definite-causal ordered  and causal-superposition strategies, but can be attained with the aid of  \textbf{TF}-assistance. We refer to this phenomenon as ``metrological activation'', establishing indefinite time direction as a genuine resource for overcoming fundamental limitations of noisy quantum metrology.

We also exhibit that this metrological activation have other implications that extend  beyond precision sensing. The quantum time-flip has been shown to improve the performance of various quantum information processing tasks, including channel discrimination~\cite{chiribella2022quantum}, quantum communication~\cite{Liu2023Apr}, and quantum memory effects~\cite{karpat2024memory}, with several of these advantages supported by experimental demonstrations~\cite{guo2024experimental,stromberg2024experimental, Xia2024Jul}. At a foundational level, the possibility of performing TF raises intriguing questions concerning the thermodynamic arrow of time~\cite{Rubino2021Nov}. These developments naturally lead to the fundamental question of whether the TF supermap can be simulated within the framework of quantum circuits and a finite number of queries to the input CPTP map. Interestingly,
in the context of other higher-order quantum transformations, it has been shown that it is possible to construct a transposition and inversion supermaps for unitaries with just four queries~\cite{Yoshida2023Sep,Grinko2024Dec,Odake2024May}. For deterministic and exact simulability of a quantum switch with unitary inputs, two queries of the input unitaries suffice~\cite{chiribella2013quantum} while simulating the switch for general channels requires at least an exponential number of queries~\cite{Kristjansson2024Sep}. By exploiting metrological activation as an operational tool, we report that neither the quantum time-flip nor the transpose supermap, when acting on general CPTP maps, can be simulated deterministically and exactly by quantum combs or causal-superposition strategies with any finite number of queries.

The paper is arranged as follows. In Sec. \ref{sec:optimal_metrology}, we discuss the known results and methodologies on optimal metrology that we will be using in our work. 
In Sec. \ref{sec:Met_finite}, we provide rigorous definitions of ITDM and TF-assisted protocols and present their advantages to optimal metrology in finite usage scenario.
In Sec. \ref{sec:asympt}, we present asymptotic advantage of TF-assisted protocols over adaptive and causal superpositions.  The metrological activation of Heisenberg scaling for certain noise models with the help of time-flip operation and its unsimulability are reported in Sec. \ref{sec:unsiml}.  Our findings are summarized in Sec. \ref{sec:conclu}.

\section{Framework for optimal quantum Metrology}
\label{sec:optimal_metrology}

Given $N$-queries of a completely positive and trace preserving map, $A_\theta:\mathcal{L}(\mathcal{H})\to\mathcal{L}(\mathcal{K})$, the metrological protocol of estimating $\theta$ consists of three steps -- $(1)$ prepare the probe state $\rho$, $(2)$ evolve the probe state via $N$-queries of $ A_\theta$ along with control operation, error correction, feedback etc. to obtain the encoded state $\rho_\theta$, and $(3)$ perform decoding measurement on $\rho_\theta$ in the basis of symmetric logarithmic derivative (SLD) \cite{Braunstein1994May, paris2009quantum} to obtain quantum Fisher information (QFI), $\mathcal J(\rho_\theta)$. In local metrology, the achievable precision of the estimate, quantified by the variance, $\Delta \theta = \langle \theta^2 \rangle - \langle \theta \rangle^2$ is bounded below by the inverse of QFI, i.e., $(\Delta \theta)^2 \geq 1/[M \mathcal{J}(\rho_\theta)]$ known as the quantum Cramer-Rao bound  \cite{helstrom1969quantum, Giovannetti2011Apr}, where $M$ is the number of repititions of the experiment. Hence, one way to increase the precision is to find optimal higher order quantum maps, termed as strategy sets \cite{liu2023optimal}, that convert a bunch of parametrized channels to the encoded state in order to maximize the QFI. Exact calculation of QFI can be translated to semidefinite programming (SDP) algorithms via maximization over purification (MOP) and iterative see-saw methods (ISS) \cite{Macieszczak2013Dec, Toth2018Jan, Demkowicz-Dobrzanski2011Jun, Macieszczak2014Oct, liu2023optimal, Fujiwara2008May, Demkowicz-Dobrzanski2012Sep, Demkowicz-Dobrzanski2014Dec, Chiribella2012Dec, Altherr2021Aug} (further details in Ref.~\cite{Liu2024Dec, Kurdzialek2025Jan} and a python package in Ref.~\cite{Dulian2025Jun}).

\begin{figure}[h] 
    \centering
\includegraphics[width=0.45\textwidth]{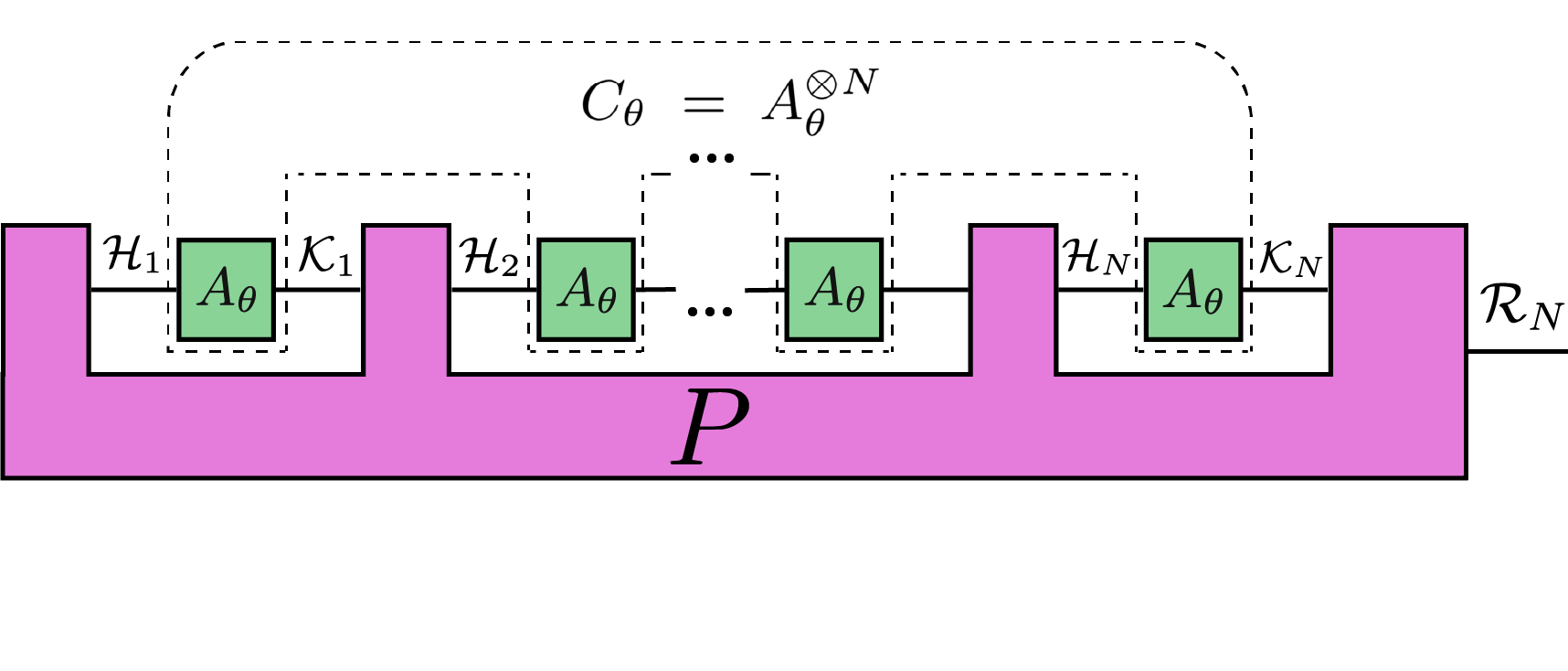}
        \caption{\textbf{ Circuit diagram for a general metrological protocol.} $N$ channels $A_\theta$ are composed with a strategy $P$ to obtain a parameter-dependent state $\rho_\theta$ upon which measurement is to be performed. }
    \label{fig:mop}
\end{figure}

\textit{MOP. } Let the supermap which transforms $C_\theta=A_\theta^{\otimes N}$ into $\rho_\theta^N\in \mathcal R_N$ be $P:\mathcal L(\otimes_{i=1}^{N}\mathcal K_i)\to \mathcal L(\otimes_{i=1}^N\mathcal H_i\otimes \mathcal R_N)$  (see Fig.~\ref{fig:mop}). Now, given any quantum operation $X:\mathcal{L}(\mathcal{H}_\textnormal{in})\to\mathcal{L}(\mathcal{H}_\textnormal{out}),$ \footnote{Quantum operations are denoted by Times New Roman font with italics, while CJ operators are denoted by Sans Serif font.} the corresponding Choi-Jamiołkowski (CJ) operator, i.e., $ \mathsf{X}=X\otimes\mathcal I_{\mathcal H_{in}}(\ketbra{\mathbb 1})$ where $\ket{\mathbb 1}=\sum_{j}\ket{j}_{\mathcal{H}_{in}}\ket{j}_{\mathcal{H}_{in}}$ is the unnormalized maximally entangled state in $\mathcal H_{in}\otimes\mathcal{H}_{in}$. Therefore, the CJ matrix of $P$ is given by $\mathsf P \in \mathcal L(\bigotimes_{i=1}^N(\mathcal H_i\otimes \mathcal K_i)\otimes \mathcal R_N)$. If $A_\theta(*)=\sum_{i=1}^rK_{i,\theta}(*)K_{i,\theta}^\dagger$ with $\sum_{i=1}^r K_{i,\theta}^\dagger K_{i,\theta}=\mathbb I$ and $r$ being the rank of the Kraus representation, we define $\mathsf C_\theta = \mathsf A_\theta^{\otimes N} = (\sum_{i=1}^{r}| {\mathsf K}_{i,\theta} \rangle \langle {\mathsf K}_{i,\theta}|)^{\otimes N}$. Note that, although the decomposition of $A_\theta$ is not unique, choosing an arbitrary decomposition suffices. Now, given a set of strategies $\textbf{Strat}\equiv\{\mathsf P|\mathsf P\geq 0, \text{rank}(\mathsf P)=1, \text{Tr}_{\mathcal R_N}\mathsf P=\widetilde{\mathsf P}\in\widetilde{\textbf{Strat}}\}$, the maximum QFI can be written as (see Appendix~\ref{sec:MOP})
\begin{eqnarray}
    \nonumber \mathcal J^{{\textbf{Strat}}}(\mathsf C_\theta)=\max_{\mathsf P\in{\textbf{Strat}}}\mathcal J(\mathsf C_\theta\star\mathsf P)=4\min_h\max_{\widetilde P\in \widetilde{{\textbf{Strat}}}}\text{Tr}[{\Omega}(h)\widetilde{\mathsf P}],\\
    \label{eq:def_mop}
\end{eqnarray}
where ${\Omega}(h)=\big(\sum_i| \dot{\mathsf K}_{i,\theta}(h) \rangle \langle \dot{\mathsf K}_{i,\theta}(h)|^T\big)^{\otimes N}$ which satisfies $\dot{\mathsf K}_{i,\theta}(h) = \dot{\mathsf K}_{i,\theta} -\iota\sum_j h_{ij} \mathsf K_{\theta, j}$ and $h$ is an $r$-dimensional arbitrary Hermitian matrix for some fixed Kraus representation corresponding to ${\mathsf K}_{i,\theta}$. Here `$\star$' is the link product defined for two arbitrary operators, $V\in\mathcal L(\mathcal A_1\otimes\mathcal A_C)$ and $W\in\mathcal L(\mathcal A_C\otimes\mathcal A_2)$ as $V\star W=\text{Tr}_{\mathcal A_C}[(V\otimes \mathbb I_{\mathcal A_2})(\mathbb I_{\mathcal A_1}\otimes W^{T_{\mathcal A_C}})],$ with $T_{\mathcal A_C}$ being the transpose operation with respect to $\mathcal A_C$.

Eq.~\eqref{eq:def_mop} can now be transformed into semidefinite programming, which utilizes the dual space of these strategy sets. In particular, there should exist the affine spaces of Hermitian operators $\{s^i\}_{i=1}^n$ such that 
\begin{eqnarray}
    \widetilde{\textbf{Strat}}\equiv\operatorname{Convex Hull}[\cup_{i=1}^n\{s^i\geq 0|s^i\in S^i\}].
\end{eqnarray}
In the literature, the existing strategies consist of those with definite causal order, such as parallel (\textbf{Par}) and adaptive (\textbf{AD}) strategies. Now, \textbf{Par} corresponds to the parallel usage of $N$ channels, $A_\theta$ along with auxiliary systems (see Fig.~\ref{fig:Par_AD_CS}(a)). On the other hand, the most general causally ordered strategy is the \textbf{AD} strategy, where $N$ channels are used sequentially, aided by auxiliary system. In Fig.~\ref{fig:Par_AD_CS}(b), it is evident that only the output of the previous channel can affect the input of the next channel, together with control unitary between each of the two channel usages. We also consider the causal superposition (\textbf{CS}) strategies \cite{chiribella2013quantum, Liu2024Dec}, see Fig.~\ref{fig:Par_AD_CS}(c), where the channels are probed in superposition of definite causal orders. Lastly, the most general indefinite causal ordered (\textbf{ICO}) strategies are those which cannot be written as convex combinations of definite causal order processes (combs). See Appendix~\ref{sec:appendix_strategies} for mathematical definition of the strategy sets, $  \textbf{Strat} \in \{\textbf{Par, AD, CS, ICO} \}$.

Now, $\mathcal{J}^{\textbf{Strat}}(\mathsf C_\theta)$ can be cast into an SDP as 
\begin{equation}
    \begin{aligned}
     \mathcal{J}^{\textbf{Strat}}(\mathsf C_\theta) =& 4\min_{\lambda,\mathcal{O}^i,h} \lambda, \notag \\
    \mathrm{s.t.}\ &\,A^i \ge 0,\ \mathcal{O}^i\in \bar S^i,\ i=1,\dots,n,
    \end{aligned}  
\end{equation}
where
 \begin{equation} 
   A^i =  \begin{pmatrix}
        \lambda I & B^\dagger \\
        B & \mathcal{O}^i
        \end{pmatrix},
\end{equation}
with    
\begin{eqnarray}
B = \left(|{\dot{\mathsf K}_{1,\theta}(h)}\rangle~\cdots~|{\dot{\mathsf K}_{r,\theta}(h)}\rangle\right),
\end{eqnarray}
and $I$ is an $r\times r$ dimensional identity matrix. Here, $\bar S^i$ is the dual affine space of  $S^i$, i.e., $\bar S^i=\{\mathcal O|\mathcal O^\dagger=\mathcal O, ~\Tr[\mathcal Os] = 1, ~ s \in S^i\}$.

Although extremely effective, MOP scales poorly with the increasing number of channel usages due to the fact that auxiliary dimension cannot be controlled here. Typically, it becomes impossible to go beyond $N=4$. Thus, alternative algorithms which
could scale to higher usage of channels are desirable.  Here we consider tensor network adaptation of the iterative see-saw method \cite{Kurdzialek2025Jan}. While ISS is advantageous in its own right since, unlike MOP, it allows for optimizations of other loss
functions such as Bayesian quadratic costs, enabling optimizations for Bayesian metrology. For the purpose
of our current work, the main advantage of the ISS method comes from its ability to be moulded for adaptive metrology by fixing the dimensions of the auxiliary systems. This enables a tensor network approach which allows for QFI estimations for high number of channel uses (see Appendix~\ref{sec:ISS}).
While the ISS method, when coupled with the tensor network approach, allows for a significant improvement in the optimizations, it still cannot ensure that the advantages bestowed by one particular strategy set over another at some finite usage can remain as the channel usage increases. Furthermore, realistically, the optimization can only be done for small auxiliary dimensions $d_{\mathcal{A}} = 2,4$ and hence a general statement about optimizations over all combs cannot be made.

In the following, let us now introduce the quantum time-flip and characterize the metrological protocols with its assistance.

\subsection{Characterizing time-flip-assisted metrology}
\label{sec:Met_finite}




The quantum time-flip is defined \cite{chiribella2022quantum} as a supermap which takes an input as a channel $A_\theta$ and outputs the controlled channel $T_{A_\theta}$. $T_{A_\theta}$ acts as $A_\theta$ when the control qubit is initialized in $\ket 0$, while $\Omega(A_\theta)$ when the control qubit is in $\ket 1$, where $\Omega$ is the input-output inversion operator. This map $\Omega$ should have certain properties including ``order reversing'', ``identity preserving'', $\Omega(\mathcal{A}) \neq \Omega(\mathcal{B})$ when channels $\mathcal A, \mathcal B$  are not identical, and $\Omega\big(q \mathcal{A} + (1-q)\mathcal{B}\big) = q\Omega(\mathcal{A}) + (1-q)\Omega(\mathcal{B})$ with $\  1\geq  q \geq 0$. Moreover, the Kraus operators $\{K_{i,\theta}\}$ of ${A_\theta}$ should satisfy bistochasticity conditions, i.e., $\sum_i K^\dagger_{i,\theta}K_{i,\theta}=\sum_i K_{i,\theta}K^\dagger_{i,\theta}=I$, with $I$ being the identity matrix on the Hilbert space of the system for time-flip operation. The operation $\Omega$ can be described as $\Omega(A_\theta)(*)=\sum_i \omega(K_{i,\theta})(*)\omega(K_{i,\theta})^\dagger$ where $\omega:K_{i,\theta}\to \omega(K_{i,\theta})$ is equivalent to either adjoint or transpose. For the purpose of this work, (see also Ref.~\cite{ Liu2023Apr, guo2024experimental, stromberg2024experimental,Agrawal2025Jul}), we choose the transpose operation to be the input-output inversion. Hence, the action of $T_{A_\theta}$ is described by  \(T_{A_\theta}(\rho_c \otimes \rho) = \sum_{i} T_{i}^{A_\theta}(\rho_c \otimes \rho) T_{i}^{\mathcal A_\theta\dagger},\) where
$T_{i}^{A_\theta} = \ket{0}\bra{0} \otimes K_{i,\theta} + \ket{1}\bra{1} \otimes K_{i,\theta}^T$ are its Kraus operators, and the first register functions as the control for the time direction.

Now, given $N$ usage of $T_{A_\theta}$, its CJ operator is defined as $\mathsf T_{\mathsf A_\theta}^{\otimes N}$. Therefore, we have the following definition:
\begin{definition}
Given $N$ queries of the channel $A_\theta$ and a strategy $P\in \textnormal{\textbf{Strat}}$, the \textnormal{\textbf{TF}}-assisted metrology is defined as $\mathcal J(\mathsf T_{\mathsf A_\theta}^{\otimes N} \star \mathsf P)$. The name \textnormal{\textbf{TF}}-assisted metrology comes from the fact that the time-flip assists the strategy $\mathsf P$. Here, given a strategy set \textnormal{\textbf{Strat}}, the maximum achievable QFI can be calculated as $\mathcal J^\textnormal{\textbf{flip\_Strat}}\equiv\mathcal J^{\textnormal{\textbf{Strat}}}({\mathsf T_{\mathsf A_\theta}^{\otimes N}})=\max_{\mathsf P}\mathcal J(\mathsf T_{\mathsf A_\theta}^{\otimes N}\star \mathsf P),$ where the maximization is performed over all strategies, $P$. 
\label{def:tfassisted2}
\end{definition}
Combining the assistance of time-flip along with \textbf{Strat}, we name the metrological process to be  $\textbf{flip}\_\textbf{Strat}$ where the strategy can be arbitrary, but the channels are time-flipped. To establish \textit{genuine metrological advantage of time-flip (GMATF)}, we also define $\textbf{rev\_Strat}$, where the assistance is given by a time reversal (\textbf{R}) operation $\Omega(\star)$, assumed to be the transpose operation in our work. This means the parametrized channel in the case of  \textbf{rev\_Strat} is given by $\Omega(A_\theta)^{\otimes N}={A_\theta^T}^{\otimes N},$ whose CJ matrix is $\mathsf{A^T_\theta}^{\otimes N}$ with $\mathsf{A^T_\theta}=\sum_{i,j=0}^{\dim(\mathcal H)-1}A_\theta^T(\ketbra{i}{j})\otimes \ketbra{i}{j}$. Consequently, we have the following:
\begin{definition}
Given $N$ queries of the channel $A_\theta$, input-output reversal metrology, termed as \textnormal{\textbf{R}}-assisted metrology or \textnormal{\textbf{rev}} metrology, where the input-output inversion operator assists some strategy set \textnormal{\textbf{Strat}} which is defined as $\mathcal{J}(\mathsf{A^T_\theta}^{\otimes N} \star \mathsf P)$ with $P\in \textnormal{\textbf{Strat}}$. In this scenario, the maximum QFI for \textnormal{\textbf{Strat}} is $\mathcal J^\textnormal{\textbf{rev\_Strat}}\equiv\mathcal J^{\textnormal{\textbf{Strat}}}({\mathsf A_\theta ^{\mathsf T}}^{\otimes N})=\max_\mathsf{P}({\mathsf A_\theta ^{\mathsf T}}^{\otimes N} \star \mathsf P)$.
\label{def:IO_inverted}
\end{definition}
Therefore, qualitatively, for a given \textbf{Strat}, we have GMATF if $\mathcal J^{\textnormal{\textbf{Strat}}}({\mathsf T_{\mathsf A_\theta}^{\otimes N}})- \max \left[\mathcal J^{\textnormal{\textbf{Strat}}}({{\mathsf{\mathsf A_\theta}}^{\otimes N}}),~\mathcal J^{\textnormal{\textbf{Strat}}}({{\mathsf A_\theta^{\mathsf T}}^{\otimes N}})\right]>0$. For completeness, we define the indefinite time directed metrology (ITDM) {corresponding to most general bidirectional quantum processes \cite{Apadula_2026}} as follows:
\begin{definition}
Given $N$ queries of the channel $A_\theta$, \textnormal{ITDM} consists of all metrological protocols which evaluate $\mathcal{J}( \mathsf A_\theta^{\otimes N} \star \mathsf P)$ such that $P\in \textnormal{\textbf{ITD}}$. Here, the strategy set $\textnormal{\textbf{Strat}} = \textnormal{\textbf{ITD}}$ contains all quantum supermaps which can take as input bistochastic non-signalling channels and output a quantum state \cite{chiribella2022quantum,Apadula_2026}.
\label{def:itdm}
\end{definition}
Since quantum combs and indefinite causal order strategies also take input bistochastic non-signalling channels to  CPTP maps, we have \textbf{ICO$\subset$ ITD}. However, without going into the details of ITDM, we concentrate on the \textbf{TF}-assisted metrology. Finally, for comparison, we denote the regular (conventional) metrology process defined in Eq.~\eqref{eq:def_mop} as $\textbf{reg\_Strat}$, i.e., $\mathcal J^\textbf{reg\_Strat}\equiv\mathcal J^\textbf{Strat}(\mathsf{A_\theta}^{\otimes N})$ which do not have any assistance of time-flip or input-output inversion. It is clear that \textbf{TF}-assisted metrology is a subset of ITDM but does not come under any definite time directed regular quantum metrological protocols, i.e., \textbf{reg\_Strat}.

In this work, we will consider \textbf{TF}-assisted metrology so as to utilize the existing strategy sets, $\textbf{Strat}\in\{\textbf{Par, AD, CS, ICO}\}$ in the literature. Therefore, we have sets, $\textbf{reg\_Strat} \subseteq \textbf{flip\_Strat} \subseteq \textbf{ITD}$ and $\textbf{rev\_Strat} \subseteq \textbf{flip\_Strat} \subseteq \textbf{ITD}$. 
\begin{figure}[h] 
    \centering
\includegraphics[width=0.5\textwidth]{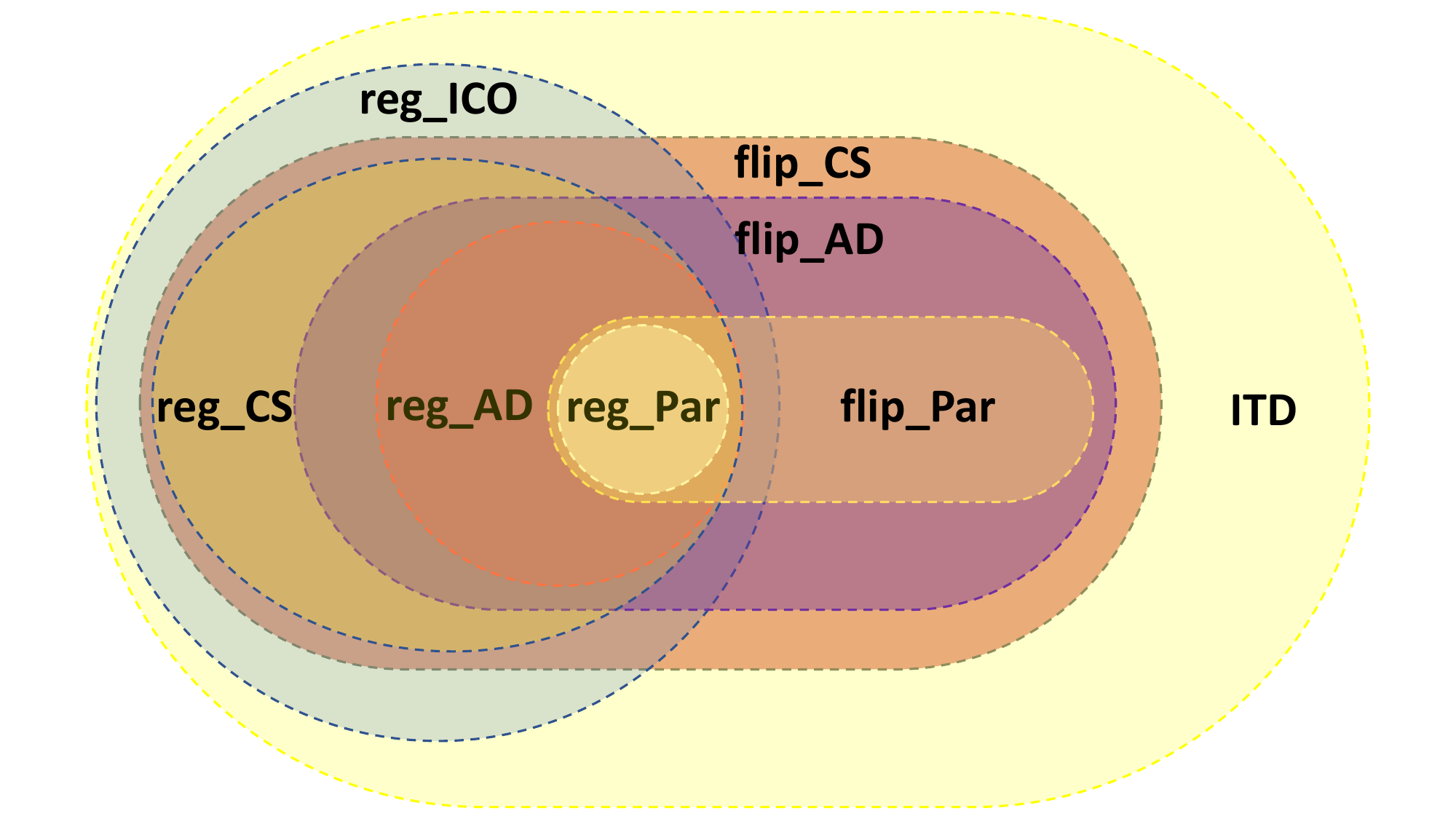}
    \caption{{ Schematic of strategies and their inclusions. All the sets are defined in the main text.}} 
    \label{fig:inclusions}
\end{figure}
See Fig. \ref{fig:inclusions} for the relations between all existing metrological protocols along with those defined in this work. Since the optimal QFI function involves optimization over all elements of a strategy set, any strategy set $A$ contained in another set $B$ cannot provide a higher optimal QFI than $B$. Furthermore, since a time-flipped channel can simply simulate a regular and input-output inverted channel by manipulating the control state, we can predict a hierarchy $\mathcal{J}^\textbf{reg\_Strat} \leq \mathcal{J}^\textbf{flip\_Strat} \leq \mathcal{J}^\textbf{ITD}$ and $\mathcal{J}^\textbf{rev\_Strat} \leq \mathcal{J}^\textbf{flip\_Strat} \leq \mathcal{J}^\textbf{ITD},$ with a crucial question being when these inequalities are strict. 

\subsection{Numerical results for finite usage of channels}
In noisy quantum metrology, where each subsystem of the probe state is affected by independent noise, we choose three prototypical noise models, namely -- $(1)$ depolarizing channel (DPC), $A^{DPC}$, $(2)$ phase damping channel (PDC), $A^{PDC}$, and $(3)$ nuclear magnetic resonance noise (NMR), $A^{NMR}$. In such a scenario, the aim is to estimate the phase $\theta$ of the unitary operator $U_\theta = e^{i\theta\vec n \cdot\vec\sigma/2}$. We take $\theta=\pi/2$ for illustrations. Our objective is to compare \textbf{TF}-assisted metrological schemes with definite causal order (DCO), specifically, the \textbf{Par} and \textbf{AD} strategies, against their corresponding unassisted counterparts as well as the \textbf{reg\_ICO} metrology.

\subsubsection{$N=3$ queries of channels}
\label{subsubsec:n=3}
Let us present the numerical results of QFI, for $N=3$. For \textbf{reg} and \textbf{rev} metrology, we calculate the optimal QFI via MOP method. However, due to increasing complexity, we can only compute the QFI of \textbf{TF}-assisted protocols via tensor network ISS method by fixing the auxiliary dimension $d_\mathcal{A}$, matrix product state (MPS) and symmetric logarithmic derivative (SLD) bond dimension to be equal to two.  Note that any increase in these dimensions to perform a more ideal comparisons may only serve to further consolidate the advantages of \textbf{TF}-assisted protocols.

\begin{figure*}
    \centering
\includegraphics[width=1\textwidth]{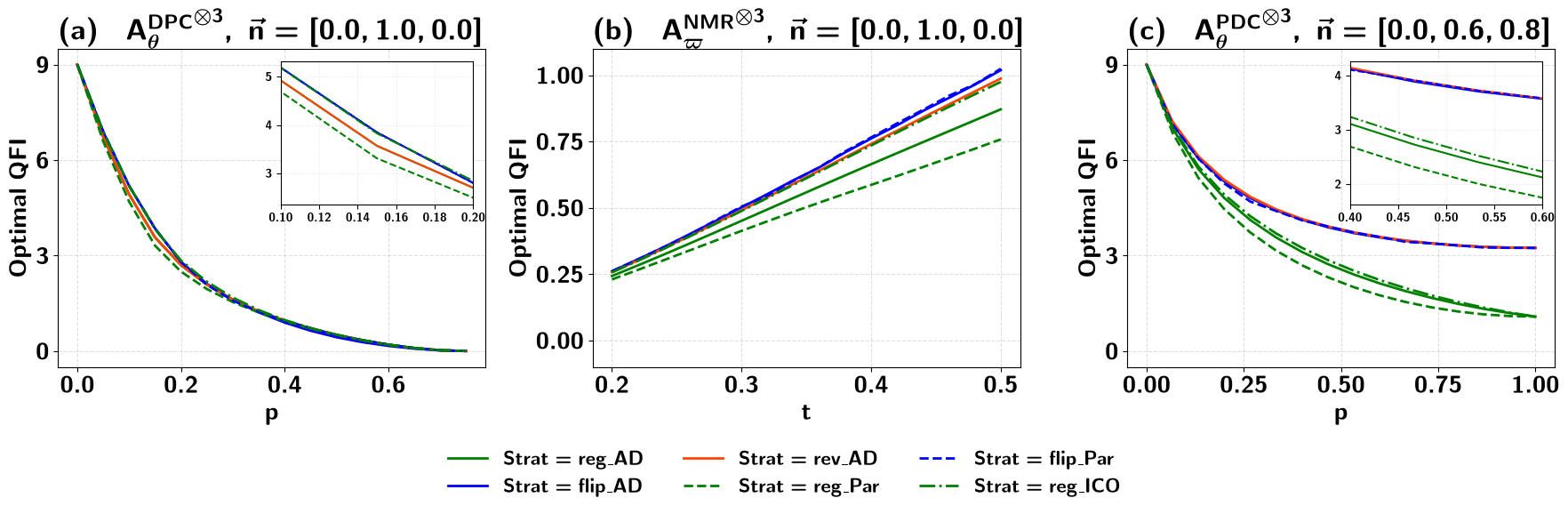}
    \caption{\textbf{ Optimal QFI  (ordinate) vs noise parameters (p) (abscissa) for various strategies and channels with finite usage $N = 3.$} (a) For a channel affected by depolarizing noise $A^{DPC}_\theta = \mathcal{U}_\theta \circ A^{DPC}$, (b) by the NMR noise $A^{NMR}_\varpi = \mathcal{U}_\varpi \circ A^{NMR}$  and (c)  by the phase damping noise $A^{PDC}_\theta = \mathcal{U}_\theta \circ A^{PDC}.$ Insets depict the variations of different strategies and channels for a particular range of \(p\) so that they can be clearly visible. All axes are dimensionless.}  \label{fig:finite_all_param}
\end{figure*}

\noindent\textit{Depolarizing noise.} The Kraus operators of $A^{DPC}$ can be written as $ \{K_0^{DPC}=\sqrt{1-p}I,~ K_1^{DPC}=\sqrt{p/3}\sigma_x,~ K_2^{DPC}=\sqrt{p/3}\sigma_y,~ K_3^{DPC}=\sqrt{p/3}\sigma_z\}$. Therefore, the effective channel, a composition of DPC and $U_\theta$ acting on the probe state, is given by $A_\theta^{DPC}(*)=\sum_{i=0}^3K_{i,\theta}^{DPC}(*){K_{i,\theta}^{DPC}}^{\dagger}$ with $K_{i,\theta}^{DPC}=U_\theta K^{DPC}_i$. Since DPC is $U(2)$-covariant channel, it follows that   $(\mathcal U_\theta\circ A^{DPC})^T=\mathcal U_\theta^T\circ A^{DPC}$ with $\mathcal U_\theta(*)=U_\theta(*)U_\theta^\dagger$. Furthermore, owing to the symmetry of the DPC about the maximally mixed state, we have $\textbf{rev\_AD}\equiv\textbf{reg\_AD}$.

In the time-flipped setup, $A_\theta^{DPC}$ undergoes the quantum time-flip supermap to give the time-flipped parametrized channel, $T_{A_\theta^{DPC}}$,  which then acts on the probe. Fixing $N=3$ and $\vec n=\hat y$, we analyze QFI by varying noise strength $p\in[0,1]$ (see Fig. \ref{fig:finite_all_param}(a)). First of all, we have $\textbf{flip\_Par}\equiv \textbf{flip\_AD}$ which is true with an maximum error tolerance of $\mathcal O(10^{-2}).$  In the region $(0.1\lesssim p\lesssim 0.2),$ we find that the assistance of time-flip outperforms its unassisted counterparts of strategies with DCO, i.e., $\mathcal J^{\textbf{AD}}({\mathsf T_{\mathsf A_\theta^{DPC}}^{\otimes 3}})=\mathcal J^{\textbf{Par}}({\mathsf T_{\mathsf A_\theta^{DPC}}^{\otimes 3}})>\mathcal J^{\textbf{AD}}({{{\mathsf A_\theta^{DPC}}}^{\otimes 3}})>\mathcal J^{\textbf{Par}}({{{\mathsf A_\theta^{DPC}}}^{\otimes 3}})$. On the other hand, although QFI of \textbf{reg\_ICO} matches with \textbf{flip\_AD}, in the range of $(0.2\lesssim p\lesssim0.6),$ we have $\mathcal J^{\textbf{ICO}}({{\mathsf A_\theta^{DPC}}^{\otimes 3}})>\mathcal J^{\textbf{AD}}({\mathsf T_{\mathsf A_\theta^{DPC}}^{\otimes 3}})$ with a maximum difference of $\mathcal O(10^{-1})$, which indicates that they are incomparable and require $d_\mathcal{A}>2$ for any conclusion. \\

\begin{figure}
    \centering
\includegraphics[width=0.45\textwidth]{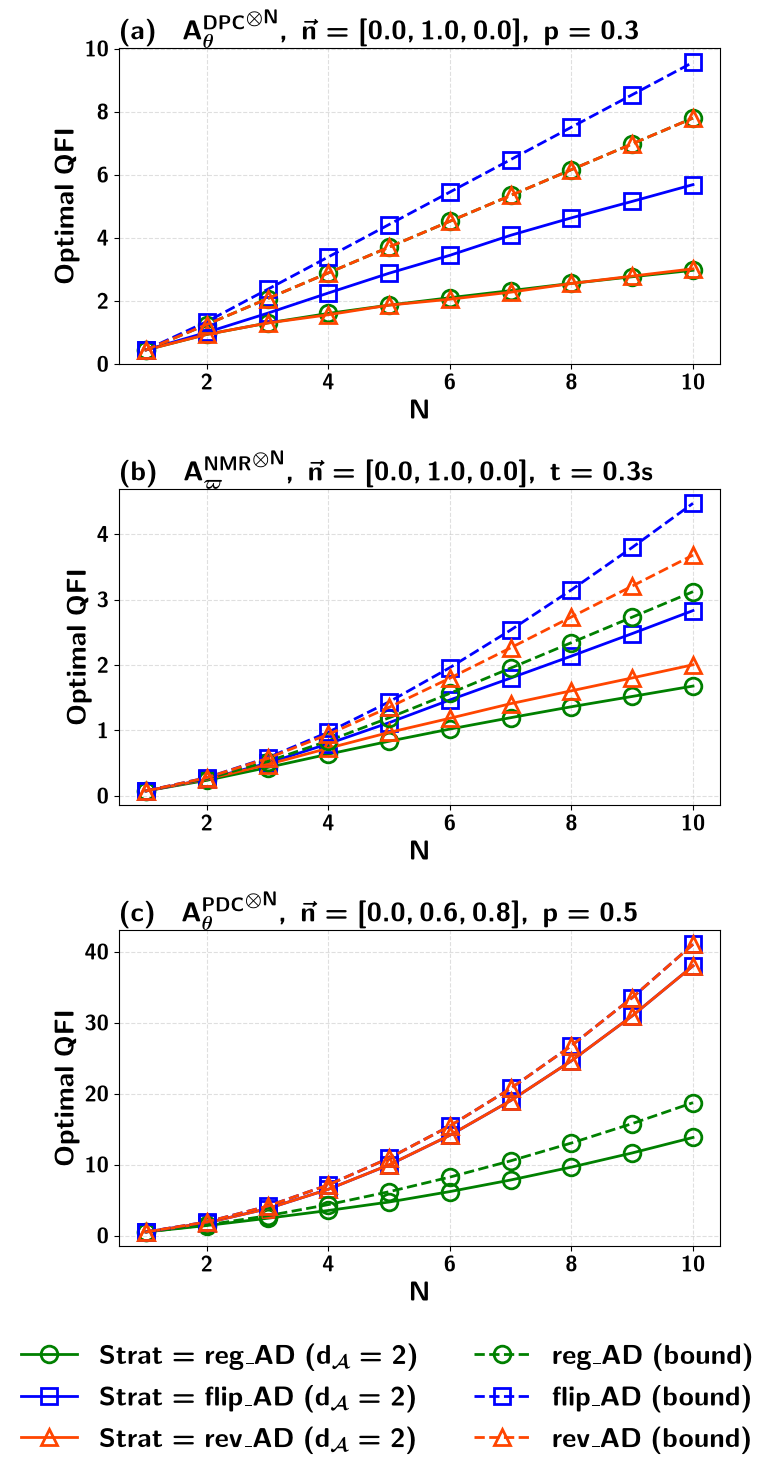}
    \caption{\textbf{ Optimal QFI  with fixed ancillary dimensions and their upper bounds (ordinate) vs usage (abscissa) for various strategies and channels with fixed parameter values.}  For channel affected by (a) depolarizing noise $A^{DPC}_\theta = \mathcal{U}_\theta \circ A^{DPC}$ (b) NMR noise $A^{NMR}_\varpi = \mathcal{U}_\varpi \circ A^{NMR}$  (c) phase damping noise $A^{PDC}_\theta = \mathcal{U}_\theta \circ A^{PDC}$. All axes are dimensionless. }
    \label{fig:finite_all_N}
\end{figure}
\noindent\textit{NMR noise.} Let us now move on to a noise model for nuclear magnetic resonance (NMR) experiments. In such experiments, the decoherence effects are characterized by longitudinal relaxation time $T_1$ and transverse relaxation time $T_2$ \cite{Nielsen2012}. The NMR noise, $A^{NMR}$, can be modelled with Kraus operators as 
\begin{eqnarray}
&&K_0^{NMR}=\sqrt{1-\alpha}\ket{1}\bra{0}, K_1^{NMR}=\sqrt{1-\beta}\ket{0}\bra{1} , \notag\\
&&K_2^{NMR}=\frac{1}{\sqrt2}\sqrt{\frac{\alpha+\beta -\sqrt{\gamma^2+(\alpha-\beta)^2}}{\gamma^2 +[\alpha-\beta-\sqrt{\gamma^2+(\alpha-\beta)^2}]^2}} \notag\\
&&\times\bigg(\Big[\alpha-\beta-\sqrt{\gamma^2+(\alpha-\beta)^2}\Big]\ket{0}\bra{0} + \gamma \ket{1}\bra{1}\bigg),\notag\\
&&K_3^{NMR}=\frac{1}{\sqrt2}\sqrt{\frac{\alpha+\beta + \sqrt{\gamma^2+(\alpha-\beta)^2}}{\gamma^2 +[\alpha-\beta+\sqrt{\gamma^2+(\alpha-\beta)^2}]^2}} \notag\\
&&\times\bigg(\Big[\alpha-\beta+\sqrt{\gamma^2+(\alpha-\beta)^2}\Big]\ket{0}\bra{0} + \gamma \ket{1}\bra{1}\bigg), \notag
\end{eqnarray}
where $\alpha = (1-a_0)e^{-t/T_1} +a_0,~ \beta = a_0e^{-t/T_1}+ 1-a_0,~ \gamma = 2e^{-t/T_2}$. In accordance with typical NMR experiments \cite{Long_2022}, we take $T_1 = 3.2s, ~T_2 = 1.1s, \text{and, } a_0 = 0.5$ (making the noise bistochastic) which signifies that at $t\gg T_1,T_2$, the qubit state equilibrates to maximally mixed state. 

In this scenario, we estimate the frequency $\varpi$ of  $A_{\varpi}^{NMR}=\mathcal U_{\varpi ~t}\circ A^{NMR}$ by choosing $\vec n=\hat y$ and $\varpi=10$kHz. For three queries of $A_{\varpi},$ we observe a strict hierarchy $\textbf{flip\_AD}=\textbf{flip\_Par}>\textbf{rev\_AD}>\textbf{reg\_ICO}>\textbf{reg\_AD}>\textbf{reg\_Par}$ with significant gaps, increasing monotonically with time. While restricting $d_{\mathcal A}=2$, in the case of the \textbf{AD} strategy set, the assistance of time-flip provides GMATF as evident from Fig.~\ref{fig:finite_all_param}(b).\\\\
\noindent\textit{Phase damping noise.} When a dephasing noise $A^{DPC}$ with Kraus operators $\{K^{DPC}_0=\sqrt{1-p},K^{DPC}_1=\sqrt{p}\ketbra0,K^{DPC}_2=\sqrt p \ketbra1\}$ is present during estimation of $\theta=\pi/2$, we demonstrate that $\textbf{flip\_AD}\equiv\textbf{rev\_AD}$ (see Fig.~\ref{fig:finite_all_param}(c)). As noise increases, the advantage of \textbf{R}-assisted metrology increases and reaches maximum at $p=1$. Since \textbf{rev\_AD} strategy matches the \textbf{flip\_AD} one, the said advantage really comes from the input-output reversal, i.e., noise acting on the probe after the application of the unitary, instead of the quantum time-flip or indefiniteness in the input-output direction. While \textbf{reg} metrology respects the following hierarchy as expected $\textbf{reg\_ICO}>\textbf{reg\_AD}>\textbf{reg\_PAR}$ in the region $p\in(0,1)$. The results of the optimal QFI by means of MOP in the case of $N=2$ is discussed in Appendix~\ref{app:optimalqfin2}.

\subsubsection{Higher number of queries of channels and fixed auxiliary dimension}

For higher values of $N\geq 4$, let us calculate the values of QFI via the ISS tensor network algorithm by fixing auxiliary dimension $d_{\mathcal A}=2$  against different noise models. In this restricted scenario, for \textbf{AD} strategy set, we observe that the assistance of time-flip provides the benefit in metrology which increases monotonically with $N$ for the depolarizing noise with all $p$ values and NMR noise (as depicted in Figs.~\ref{fig:finite_all_N}(a) and (b) for a specific noise strength $p=0.3$ (DPC) and $t=0.3s$ (NMR), respectively). In the case of dephasing channel, we again obtain $\textbf{rev\_Strat}\equiv\textbf{flip\_Strat}$ (Fig.~\ref{fig:finite_all_N}(c)).

 The entire analysis strongly indicate that for all of these noise models, the optimal  \textbf{TF}-assisted protocols with restricted auxiliary dimension $d_\mathcal{A}=2$ can outperform optimal \textbf{reg} protocols with the finite usage. Moreover, the advantage of using time-flip increases with $N$. In the subsequent section, we will show that the superiority of \textbf{TF}-assistance persists even in the asymptotic limit. These results suggest that the strategies with \textbf{TF} are resourceful for quantum metrology.

\section{Genuine metrological advantage via time-flip: Upper bounds and saturable asymptotic regime}
\label{sec:asympt}

The optimal QFI are typically hard to obtain analytically once the system size, i.e., the number of usage or the number of qubits under consideration is high. For this purpose, upper bounds on optimal QFI, particularly asymptotically saturable ones are useful in two ways. Firstly, the qualitative behavior of metrological protocols at high channel usage can be obtained. Secondly, it can determine the optimality of existing protocols, since an existing scheme that provides QFI values close to the upper bounds can be regarded as optimal with a high degree of certainty. Before proceeding to the detailed analysis, we note that, in addition to the strategy classes \textbf{Par} and \textbf{AD}, we also include \textbf{CS} strategies~\cite{chiribella2013quantum, Liu2024Dec}. For notational convenience, throughout this section, the quantities associated with \textbf{TF}-assisted metrology, $\mathcal J^{\textnormal{\textbf{Strat}}}(\mathsf T_{\mathsf A_\theta}^{\otimes N})$, and \textbf{R}-assisted metrology, $\mathcal J^{\textnormal{\textbf{Strat}}}((\mathsf A_\theta^{\mathsf T})^{\otimes N})$, are denoted by $\mathcal J^{\textnormal{\textbf{flip\_Strat}}}$ and $\mathcal J^{\textnormal{\textbf{rev\_Strat}}}$, respectively, with the channel dependence omitted for brevity. Similarly, the quantity corresponding to regular metrology, $\mathcal J^{\textnormal{\textbf{Strat}}}(\mathsf A_\theta^{\otimes N})$, is denoted by $\mathcal J^{\textnormal{\textbf{reg\_Strat}}}$, encompassing the cases $\mathcal J^{\textbf{Par}}$, $\mathcal J^{\textbf{AD}}$, and $\mathcal J^{\textbf{CS}}$.

Given $N$ number of parametrized channels $C_\theta =  A_\theta^{\otimes N}$, the most optimal bound of QFI for parallel metrological schemes \cite{Demkowicz-Dobrzanski2014Dec} reads as
\begin{eqnarray}
\mathcal{J}^{\textbf{Par}} \leq \min_{\{K_{i,\theta}\}}4\Big[N\|\alpha\| + N(N-1)\|\beta\|^2\Big],
\label{eq:par}
\end{eqnarray}
where $\|A\|=\sup_{\|v\|=1}\|A(v)\|$ is the operator norm of operator $A$ with $\alpha = \sum_i\dot K_{i,\theta}^\dagger \dot K_{i,\theta}$, and $\beta = \sum_i\dot K_{i,\theta}^\dagger K_{i, \theta}$. Note that, if there exists a Kraus representation for which $\beta = 0$, only SS is achievable. Interestingly, the above bound (Eq.~\eqref{eq:par}) of QFI for $\textbf{Par}$ is asymptotically saturable \cite{Zhou_2021} both in the case of SS and in the case of $\beta\neq 0$, which correspond to Heisenberg scaling. In case of $\textbf{AD}$ and $\textbf{CS}$ strategies, the upper bound is calculated as \cite{Kurdzialek2023Aug}
\begin{eqnarray}
\mathcal{J}^{\textbf{AD}} \leq \min_{\{K_{i,\theta}\}^{\times N}} 4a_N, ~~~~  \mathcal{J}^\textbf{CS} \leq \min_{\{K_{i,\theta}\}} 4a_N,
\end{eqnarray}
where in the case of $\textbf{AD},$ we have
\begin{eqnarray}
\min_{\{K_{i,\theta}\}^{\times N}} a_N &=& \bar a_N~~\text{ with   }~~ \bar a_0 = 0, \text{ and } \notag\\
\bar a_{j+1} = \min_{\{K_{i,\theta}\}}\Big[\bar a_j &+& \|\alpha_{\{K_{i,\theta}\}}\| + 2\|\beta_{\{K_{i,\theta}\}}\|\sqrt{\bar a_j}\Big].
\label{eq:a_minimisation}
\end{eqnarray}
Here, the optimization is carried out in a recursive manner. Specifically, the minimization of 
$a_{j+1}$ over $\{K_{i,\theta}\}^{\times (j+1)}$ is achieved by substituting the minimal value 
of $a_j$ over $\{K_{i,\theta}\}^{\times j}$, denoted by $\bar{a}_j$, and by selecting an optimized Kraus representation for the 
$(j+1)$-th channel. It has been shown that both $\textbf{AD}$ and $\textbf{CS}$ are asymptotically saturable to the $\textbf{Par}$ bound, hence achievable \cite{Kurdzialek2023Aug}. 

To numerically optimize Eq.~\eqref{eq:a_minimisation}, note that for any function $f(\|\alpha_{\{K_{i,\theta}\}}\|,\|\beta_{\{K_{i,\theta}\}}\|)$ which is increasing under both the arguments, can be equivalently expressed as (see Ref. \cite{Kurdzialek2023Aug})
\begin{equation}
\min_{\{K_{i,\theta}\}}f(\|\alpha_{K_{i,\theta}}\|, \|\beta_{K_{i,\theta}}\|) = \min_{b \in [\gamma, \delta]}f(q(b), b), 
\label{eqn:kurzgiak}
\end{equation}
where 
\begin{eqnarray}
q(b) = \min_{\{K_{i,\theta}\}, ~ \|\beta_{\{K_{i,\theta}\}}\|\leq b} \|\alpha_{\{K_{i,\theta}\}}\|,
\end{eqnarray}
and $\gamma = \min_{\{K_{i, \theta}\}}\|\beta_{\{K_{i,\theta}\}}\|$, $\delta = \min_{\{K_{i, \theta}\}}\sqrt{\|\alpha_{\{K_{i,\theta}\}}\|}$. Since it suffices to consider Kraus representations of the form, $\dot{K}_{i,\theta}(h) = \dot{K}_{i,\theta} -\iota\sum_{j=1}^r h_{ij} K_{j,\theta}$, minimization over Kraus representation can be reduced to minimization over $h$. Moreover, since all the upper bounds including Eq.~\eqref{eq:a_minimisation} are indeed the functions of the form in Eq.~\eqref{eqn:kurzgiak}, the upper bounds can be efficiently calculated by formulating SDPs to find $\gamma$ and $\delta$ via the Schur's complement condition and then by calculating $f(h(b), b)$ for a large number of values of $b$ in the range $[\gamma,\delta]$ to figure out the minimum (see Ref. \cite{Kurdzialek2023Aug} for further details). 

In this premise, two natural questions arise -- $(1)$ How do the bounds of various \textbf{TF}-assisted metrological protocols behave? $(2)$ Are the \textbf{TF}-assisted bounds of $\textbf{Par, AD, CS}$ saturate to a single value, which was also true in the case of $\textbf{reg}$ protocols as discussed above?

\begin{figure*}
    \centering
\includegraphics[width=1\textwidth]{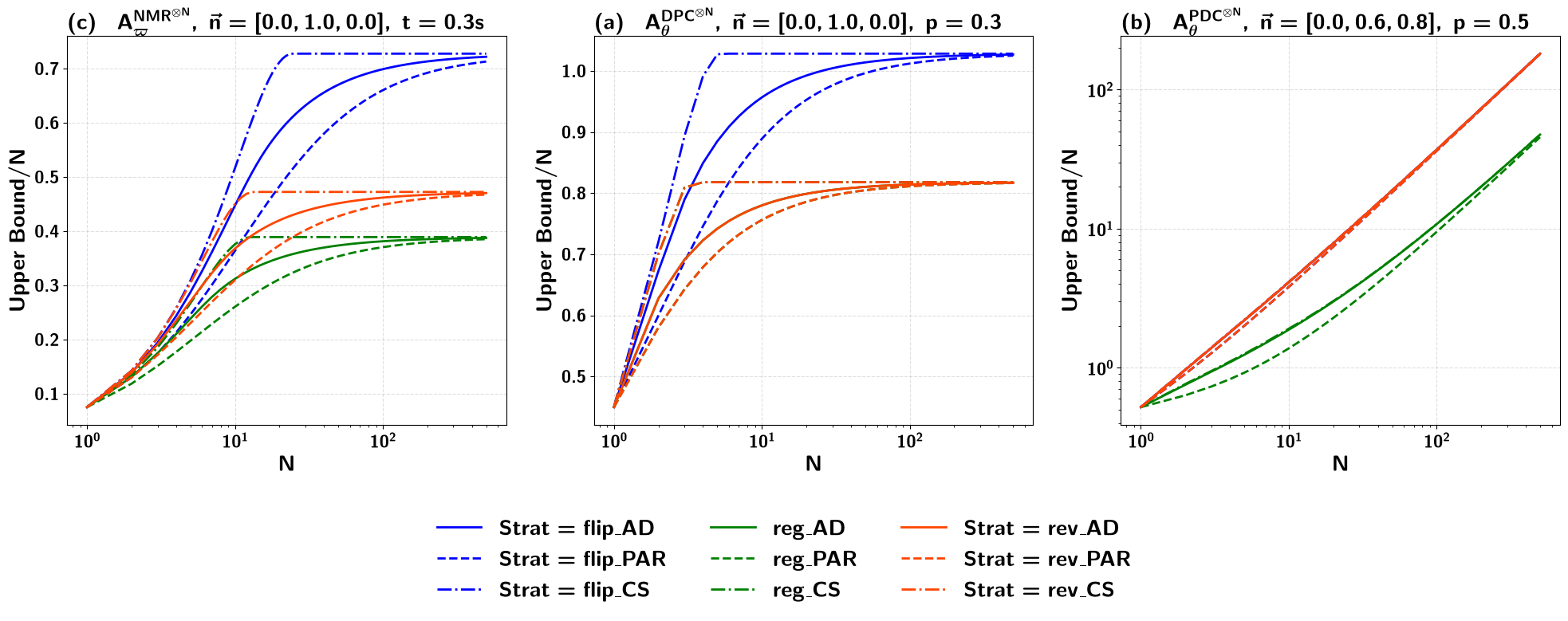}
    \caption{\textbf{Upper bounds for optimal QFI per usage of channel (ordinate) for noisy channels via various strategies against the  number of usages $N$ (abscissa).} TF-assisted strategies can outperform regular and input-output inversion strategies, breaking the asymptotic equivalence. All axes are dimensionless.}
    \label{fig:asymp_all_N}
\end{figure*}

We address these issues in the following lemma. 
\begin{lemma}
    For a given value of $b$, the function $q(b)$ for the \textbf{TF}-assisted metrology satisfies the inequality $q_{\textbf{flip}}(b) - \max\left[q_{\textbf{reg}}(b) , q_{\textbf{rev}}(b) \right] \geq 0,$ where $q_{\textbf{reg}}(b)$, $q_{\textbf{rev}}(b) $ and $q_{\textbf{flip}}(b) $ correspond to the channels $A_\theta$, $\Omega(A_\theta)$, and $T_{A_\theta}$ respectively.
    \label{lemma:g(b)lemma}
\end{lemma}

\begin{proof}
    According to the definition of $q(b)$, we have 
    \begin{eqnarray}
        &q_{\textbf{flip}}(b) = \min_{T_i^ { A_\theta},~ \|\beta_{\textbf{flip}}\| \leq b}\|\alpha_{\textbf{flip}}\|,
    \end{eqnarray}
    where $x_\textbf{flip}=\ketbra 0 \otimes x_\textbf{reg}+\ketbra 1 \otimes x_\textbf{rev}$ for $x=\alpha,\beta$ with $\alpha_{\textbf{reg}} = \sum_i \dot K_{i,\theta}^\dagger \dot K_{i,\theta}$,  $\alpha_{\textbf{rev}} = \sum_i \dot K_{i,\theta}^{T\dagger} \dot K_{i,\theta}^T$, $\beta_{\textbf{reg}} = \sum_i \dot K_{i,\theta}^\dagger K_{i,\theta}$ and  $\beta_{\textbf{rev}} = \sum_i \dot K_{i,\theta}^{T\dagger} K_{i,\theta}^T$. However, it suffices to consider Kraus representations of the form \cite{Demkowicz-Dobrzanski2012Sep},
    \begin{eqnarray}
  \dot{T}_{i}^{ A_\theta}(h) &=& \dot{T}_{i}^{ A_\theta} -\iota\sum_j h_{ji} T_{i}^{ A_\theta},\\
  &=& \ket{0}\bra{0} \otimes \Big( \dot K_{i, \theta} - \iota\sum_j h_{ji} K_{i, \theta}\Big) \notag\\
  &&+ \ket{1}\bra{1} \otimes \Big(\dot K_{i, \theta}^{T} - \iota\sum_j h_{ji}K_{i, \theta}^{T}\Big),
  \end{eqnarray}
which means   
\begin{eqnarray}
        &q_{\textbf{flip}}(b) = \min_{h,~ \|\beta_{\textbf{flip}}\| \leq b}\|\alpha_{\textbf{flip}}\|.
    \end{eqnarray}
Again, we have
    \begin{eqnarray}
        \|\alpha_\textbf{flip}\|=\max(\|\alpha_\textbf{reg}\|,\|\alpha_\textbf{rev}\|).
    \end{eqnarray}
Now, by min-max inequality of the two functions $f(x)$ and $g(x)$ having the same domain, given by
    \begin{eqnarray}
        \min_x\max(f(x),g(x))\geq\max(\min_x f(x),\min_x g(x) ),
    \end{eqnarray}
we can write 
    \begin{eqnarray}
    &&q_{\textbf{flip}}(b) = \min_{h,~ ||\beta_{\textbf{flip}}|| \leq b}||\alpha_{\textbf{flip}}|| \notag\\
    &&\nonumber \geq \max\Big(\min_{h,~ ||\beta_{\textbf{flip}}|| \leq b}||\alpha_{\textbf{reg}}||,~ \min_{h,~ ||\beta_{\textbf{flip}}|| \leq b}||\alpha_{\textbf{rev}}||\Big).\\
    \label{eq:fliph}
    \end{eqnarray}
    Since, $\|\beta_{\textbf{flip}}\| \leq b \iff \|\beta_{\textbf{reg}}\| \leq b \text{ and } \|\beta_{\textbf{rev}}\| \leq b$, we can rewrite Eq.~\eqref{eq:fliph} as
    \begin{eqnarray}
        &&q_{\textbf{flip}}(b)  \notag\\
    &&\nonumber \geq \max\Big(\min_{\substack{h,~ ||\beta_{\textbf{reg}}|| \leq b\\ ||\beta_{\textbf{rev}}|| \leq b}}||\alpha_{\textbf{reg}}||,~ \min_{\substack{h,~ ||\beta_{\textbf{reg}}|| \leq b\\ ||\beta_{\textbf{rev}}|| \leq b}}||\alpha_{\textbf{rev}}||\Big),\\
    &&\nonumber \geq \max\Big(\min_{\substack{h,~ ||\beta_{\textbf{reg}}|| \leq b}}||\alpha_{\textbf{reg}}||,~ \min_{\substack{h,~  ||\beta_{\textbf{rev}}|| \leq b}}||\alpha_{\textbf{rev}}||\Big).
    \end{eqnarray}
    This leads to the proof that $q_{\textbf{flip}}(b) - \max\left[q_{\textbf{reg}}(b) , q_{\textbf{rev}}(b) \right]\geq 0$.
\end{proof}
Therefore, it follows from Lemma~\ref {lemma:g(b)lemma} that in the case of SS, in the asymptotic region, for a given \textbf{Strat}$\in\{\textbf{Par, AD, CS}\}$ we have  $\mathcal J^{\textbf{flip\_Strat}}-\max\left[\mathcal J^{\textbf{reg\_Strat}},\mathcal J^{\textbf{rev\_Strat}}\right]=4q_\textbf{flip}(0)-4\max(q_\textbf{reg}(0),q_\textbf{rev}(0))\geq 0$. 
This leads us to define a quantitative measure which isolates the \textit{genuine metrological advantage} attributable to time-flip encoding.
\begin{corollary}
    In the asymptotic SS regime, $\text{S}_{q(0)}=q_\textnormal{\textbf{flip}}(0)-\max\left[q_\textnormal{\textbf{reg}}(0),q_\textnormal{\textbf{rev}}(0)\right]$ quantifies GMATF of the \textnormal{\textbf{TF}}-assisted metrological protocols.
    \label{cor:meas_ss}
\end{corollary}
We can also formulate such a measure of advantage of using the time-flip encoding procedure even when $b=0$ does not exist for any Kraus representation of $ A_\theta$. Following the similar argument, as given for the proof of Lemma~\ref{lemma:g(b)lemma}, $\gamma^2$ factor associated with \textbf{TF}-assistance is always greater than or equal to the larger of the corresponding factors for the \textbf{reg} and \textbf{rev} protocols.
\begin{corollary}
    In the case of asymptotic HS, $H_{\gamma^2}=\gamma^2_\textnormal{\textbf{flip}}-\max(\gamma^2_\textnormal{\textbf{reg}},\gamma^2_\textnormal{\textbf{rev}})$ serves as a quantitative measure of the GMATF associated with the \textnormal{\textbf{TF}}-assistance metrology.
    \label{cor:meas_hs}
\end{corollary}

To the aim of finding the hierarchy between \textbf{reg, rev} and \textbf{TF}-assisted metrology for arbitrary value of $N$, we establish the following lemma:
\begin{lemma}
    Given $f(x,y)$ as an increasing function of both $x$ and $y$ with non-negative values, for the set of strategies \textbf{Strat}$\in\{\textbf{Par, AD, CS}\}$, the \textbf{TF}-assisted metrology follows the relation:
    \begin{eqnarray}
    &&\min_{h}f(||\alpha_{\textbf{flip}}||, ||\beta_{\textbf{flip}}||)\geq \min_{h}f(||\alpha_{\textbf{reg}}||, ||\beta_{\textbf{reg}}||), \notag\\&& \text{ and }\notag\\
    &&\min_{h}f(||\alpha_{\textbf{flip}}||, ||\beta_{\textbf{flip}}||)\geq \min_{h}f(||\alpha_{\textbf{rev}}||, ||\beta_{\textbf{rev}}||).
    \label{eq:f_greater}
    \end{eqnarray}
    \label{lem:f_greater}
\end{lemma}
\begin{proof}
    Using Eq.~\eqref{eqn:kurzgiak}, we can write
\begin{eqnarray}
\min_{h}f(||\alpha_{\textbf{flip}}||, ||\beta_{\textbf{flip}}||) &=& \min_{b \in [\gamma_{\textbf{flip}}, \delta_{\textbf{flip}}]}f(q_{\textbf{flip}}(b), b), \notag \\
\min_{h}f(||\alpha_{\textbf{reg}}||, ||\beta_{\textbf{reg}}||) &=& \min_{b \in [\gamma_{\textbf{reg}}, \delta_{\textbf{reg}}]}f(q_{\textbf{reg}}(b), b), \notag \\
\min_{h}f(||\alpha_{\textbf{rev}}||, ||\beta_{\textbf{rev}}||) &=& \min_{b \in [\gamma_{\textbf{rev}}, \delta_{\textbf{rev}}]}f(q_{\textbf{rev}}(b), b).
\end{eqnarray}
Let $b',b'',$ and $b^*$ be the optimal values of $b$ for \textbf{reg}, \textbf{rev}, and \textbf{TF}-assisted strategies, respectively. Hence, we can write
\begin{eqnarray}
     f(q_{\textbf{reg}}(b'), b') \leq f(q_{\textbf{reg}}(b^*), b^*) \leq f(q_{\textbf{flip}}(b^*), b^*),
\end{eqnarray}
where the second inequality follows from Lemma~\ref{lemma:g(b)lemma} and the fact that $f$ is increasing in its inputs. With similar arguments, it is clear that
\begin{eqnarray}
     f(q_{\textbf{rev}}(b''), b'') \leq f(q_{\textbf{rev}}(b^*), b^*) \leq f(q_{\textbf{flip}}(b^*), b^*).
\end{eqnarray}
The above relations translate to the proof of the relation in Eq.~\eqref{eq:f_greater}.
\end{proof}

Utilizing Lemma~\ref{lem:f_greater} and Eq.~\eqref{eq:a_minimisation}, we arrive at the following theorem:
\begin{theorem}
    Given the set of strategies \textbf{Strat}$\in\{\textbf{Par, AD, CS}\}$, the \textbf{TF}-assisted, \textbf{reg} and \textbf{rev} metrology satisfy the hierarchy as
    \begin{eqnarray}
        \mathcal J^{\textbf{flip\_Strat}}\geq \mathcal J^{\textbf{reg\_Strat}} \text{, and  }\mathcal J^{\textbf{flip\_Strat}}\geq \mathcal J^{\textbf{rev\_Strat}}.
        \label{eq:stratheirarchy}
    \end{eqnarray}
\end{theorem}
The preceding analysis establishes a foundation for identifying potential advantages of the \textbf{TF}-assisted metrology in comparison with its \textbf{reg} and \textbf{rev} counterparts. Furthermore, it indicates the possibility of an asymptotic inequivalence among the \textbf{Par}, \textbf{AD}, and \textbf{CS} strategies when assisted by time-flip.

The proof of Eq.~\eqref{eq:stratheirarchy} provides a theoretical basis for the hierarchy predicted earlier based on numerics and time-flip being able to simulate regular and input-output inverted channels, thus drawing a connection between the simulability of the channels and optimal QFI. This connection will be utilized in Sec. \ref{sec:unsiml} to prove several results about simulability of supermaps. The above results also serve to construct advantage measures, allowing us to compute the gains offered by the \textbf{TF}-assisted metrology by performing SDP optimizations on the Kraus operators of the desired channels.

\subsubsection{Numerical analysis from SDP}
Let us now present some examples of the prototypical noise models considered beforehand to calculate upper bounds that are asymptotically saturable. 

\textit{NMR experiments.} Using the above approach, we numerically compute the upper bounds on the QFI for the estimation of $\varpi$ using $N$ applications of $A_{\varpi}^{\mathrm{NMR}}=\mathcal U_{\varpi t}\circ A^{\mathrm{NMR}}$ alongwith the same settings as in Sec.~\ref{subsubsec:n=3}, via SDP for the aforementioned strategies (see Fig.~\ref{fig:asymp_all_N}(a)). Our investigation demonstrates that, given a strategy, QFI satisfies the strict hierarchy, $\mathcal J^{\textbf{flip\_Strat}}_{\text{NMR}}>\mathcal J^{\textbf{rev\_Strat}}_{\text{NMR}}>\mathcal J^{\textbf{reg\_Strat}}_{\text{NMR}}$. In the asymptotic limit, the  QFI saturates to $\mathcal J^\textbf{reg\_Strat}/N\sim0.39$, $\mathcal J^\textbf{rev\_Strat}/N\sim 0.47,$ while in the case of the assistance of time-flip, $\mathcal J^\textbf{flip\_Strat}/N\sim 0.73$. Therefore, although in the cases of \textbf{reg}, \textbf{rev}, \textbf{flip}, all the considered strategies of the set \textbf{Strat} converge to the same value of QFI, the results of the above prototypical scenario enables the formulation of an important result.
\begin{proposition}
    Given any \textnormal{\textbf{Strat}}$\in\{\textnormal{\textbf{Par, AD, CS}}\}$, there exist quantum channels for which the use of \textnormal{\textbf{TF}}-assistance provides an asymptotic advantage in the estimation of the channel QFI compared to the protocol without such assistance. 
\end{proposition}
Therefore, the asymptotic equivalency \cite{Kurdzialek2023Aug} of \textbf{PAR, AD, CS} metrological protocols break down under the assistance of time-flip and input-output inversion or time reversal.  Note that it has recently been demonstrated that \textbf{TF}-assisted metrological protocols reduce the root mean squared error in certain phase estimation tasks by a constant factor of 2 \cite{guo2024experimental}. 

\textit{Depolarizing and dephasing channels.} By examining the upper bounds for the case of depolarizing channels, it is evident that TF-assisted protocols \textbf{flip\_Strat} saturate to substantially higher values of QFI per channel usage $\mathcal{J}/N\sim 1.03$ than $\textbf{reg\_Strat}\equiv \textbf{rev\_Strat}$ strategies $\mathcal{J}/N \sim 0.82$, demonstrating a genuine advantage through the assistance of time-flip (see Fig.~\ref{fig:asymp_all_N}(b) with $P_{\text{DPC}} = 0.3$ and $\vec n = \hat y$). For the phase-damping channel, it is shown that for $\vec n = \hat y,$ the assistance of \textbf{TF} achieves QFI as $N^2$ irrespective of the noise strength $p_\text{PDC}.$ Let us now choose $\vec n=[0,0.6,0.8],$ for which the  \textbf{TF}-assisted metrology is equivalent to the \textbf{rev} scheme, as shown in Fig.~\ref{fig:asymp_all_N}(c). In the asymptotic limit, the protocols for NMR and depolarizing noise achieve standard limit while, for dephasing noise, the protocol achieves HS. In all three scenarios, \textbf{TF} and \textbf{rev} provide a constant factor advantage over the \textbf{reg} scheme while retaining the same scaling.

\begin{figure}[h]
    \centering
\includegraphics[width=0.5\textwidth]{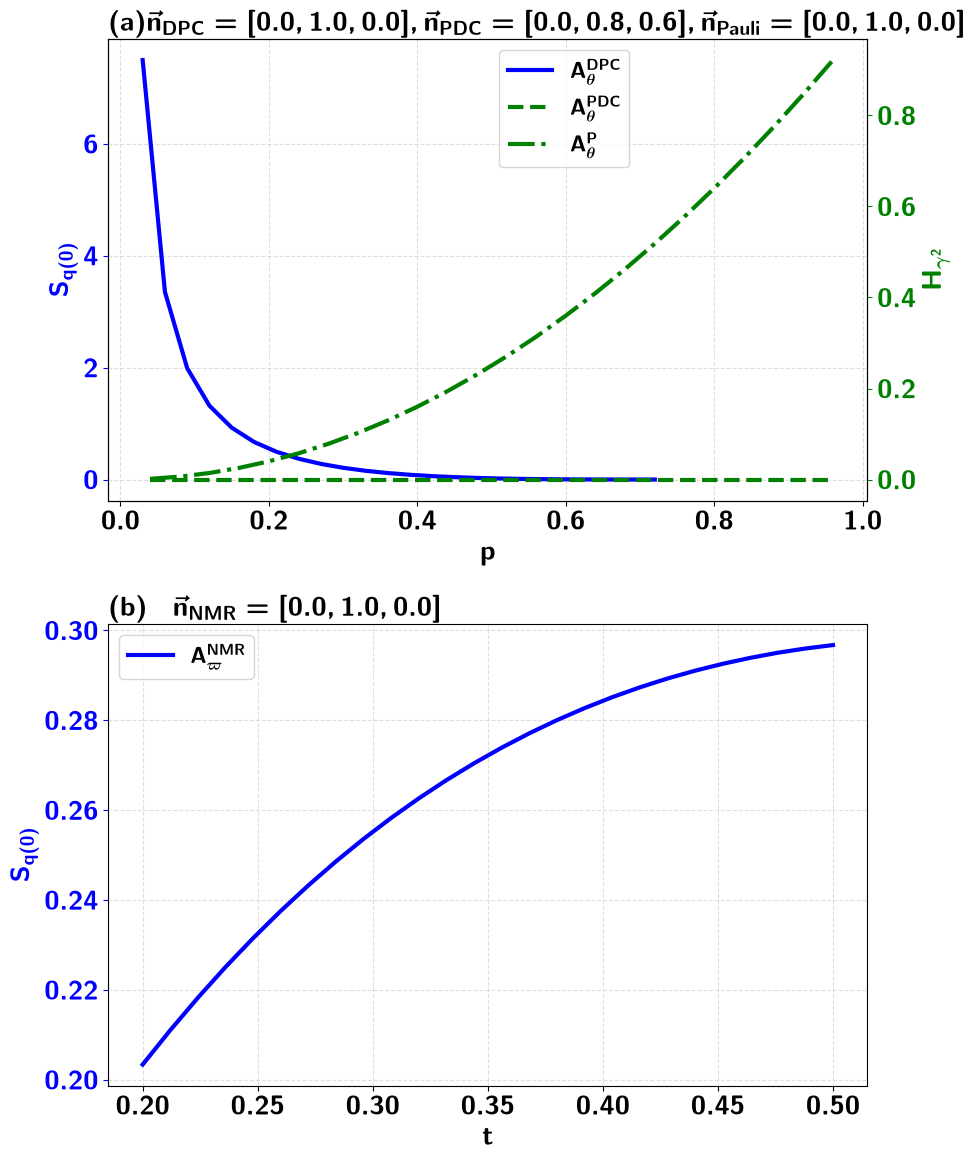}
    \caption{\textbf{Advantage measures (ordinate) with respect to noise parameter (abscissa) for different metrological tasks.} (a)  GMATF decreases rapidly for the depolarizing channel, and the metrological benefit is found for the entire range of \(p\) of a specific Pauli noise considered here,  while no benefit is found for the phase damping one. (b) Genuine metrological advantage is guaranteed for the NMR noise. All axes are dimensionless.} 
    \label{fig:advantage_meas}
\end{figure}

\textit{GMATF.} Invoking Corollaries~\ref{cor:meas_ss} and~\ref{cor:meas_hs}, we assess whether the time-flip encoding scheme offers a genuine metrological advantage across a range of noise strengths in the asymptotic channel-use limit. We compute \(S_{q(0)}\) for both the depolarizing channel and the NMR noise model (see Fig.~\ref{fig:advantage_meas}) and $H_{\gamma^2}$ for the phase damping channel with respect to noise parameters. In the NMR case, a genuine advantage is present throughout the relevant parameter regime and increases with the evolution time \(t\). In contrast, under depolarizing noise, the benefit exhibits a rapid decay as a function of noise strength, dropping below $\mathcal O(10^{-2})$ once $p \gtrsim 0.5$,  while, for the dephasing channel, there is no noise regime that yields a genuine advantage for \textbf{TF}-assisted metrology, as it is asymptotically equivalent to \textbf{rev} metrology (Fig.~\ref{fig:asymp_all_N}(c)). 

\begin{figure*}
    \centering
\includegraphics[width=1\textwidth]{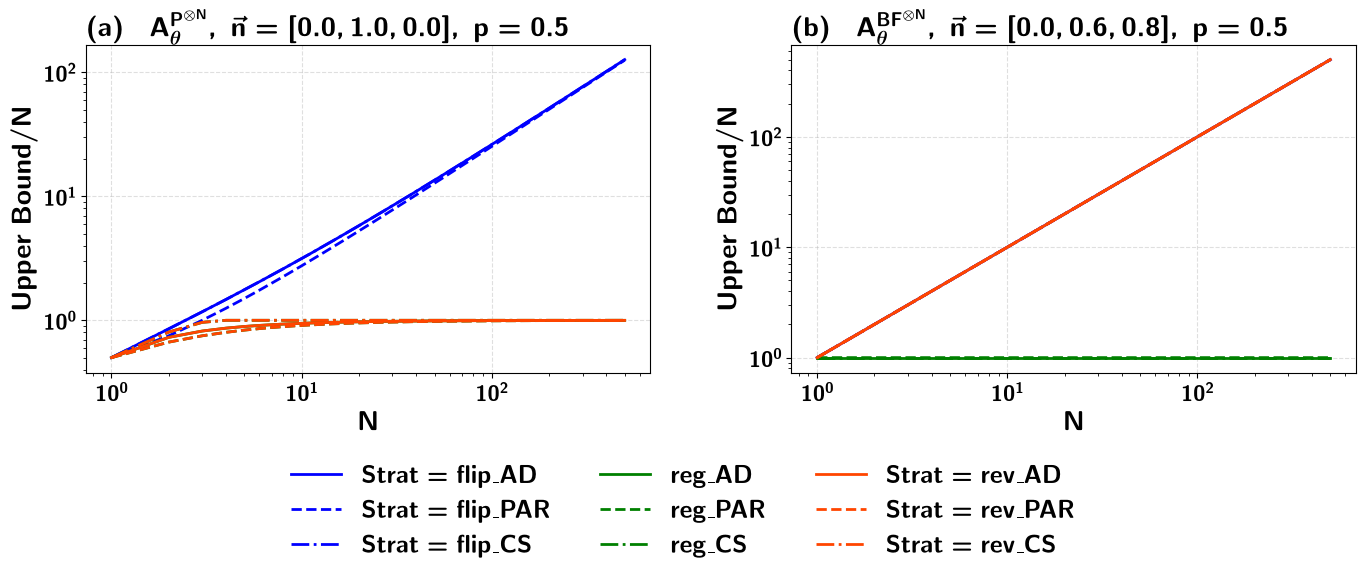}
    \caption{\textbf{Upper bounds for optimal QFI per usage of channel (ordinate) for channels affected by Pauli and bit-flip noises for various strategies with $N$ (abscissa).} TF-assisted strategies and input-output inversion provide a higher QFI scaling than the other strategies, {\bf rev} and {\bf reg}. All axes are dimensionless.  }
    \label{fig:asymp_all_N_2}
\end{figure*}

\section{Metrological activation: Unsimulability of time-flip and transpose supermaps} \label{sec:unsiml}

In the asymptotic scenario, given $N$ queries of a channel $A_\theta$ which provides no advantage, i.e., SS for \textbf{reg} metrology, if the corresponding \textbf{R}- or \textbf{TF}-assisted metrology can achieve HS, we term this phenomenon as ``metrological activation''. To demonstrate the same, consider the Pauli channel $A^P$ having Kraus representation
\begin{eqnarray}
    \nonumber K_1^P=\sqrt{p_x}\sigma_x, K_2^P=\sqrt{p_y}\sigma_y,
    K_3^P =\sqrt{p_z}\sigma_z,
\end{eqnarray}
satisfying $\sum_{i=x,y,z}p_i=1$. Specifically, let us take $p_x= p_z = \frac{1-p}{2},~ p_y=p$ where $p\in(0,1)$. Similar to the previous noisy channels, the aim here is to estimate $\theta$ of $A_\theta^P=\mathcal U_\theta\circ A^P$ with $\vec{n} = \hat{y}$. In the asymptotic regime, we find that $\mathcal{J}^{\textbf{flip\_Strat}}  \sim p_y^2N^2,$ while  $\mathcal{J}^{\textbf{reg\_Strat}} = \mathcal{J}^{\textbf{rev\_Strat}} \sim N$ (here $||\beta|| \sim 10^{-10}$) (for $p = 0.5,$ $\mathcal{J}^{\textbf{flip\_Strat}} \sim 0.25N^2$ as shown in Fig.~\ref{fig:asymp_all_N_2}(a)). This establishes the occurrence of metrological activation. The GMATF, in this case, denoted by $H_{\gamma^2}$, increases with $p$. Moreover, we simulate $10^5$ number of random $A^P$ channels satisfying $p_x,p_y,p_z\neq 0$. In all such cases, no quantum comb or causal superposition supermap could provide Heisenberg scaling but once \textbf{TF}-assisted protocol is used, a scaling of $p_y^2N^2$ is observed in the asymptotic limit. Hence, metrological activation is conferred to this specific class of channels $A^P$.

\begin{figure*}
    \centering
\includegraphics[width=1\textwidth]{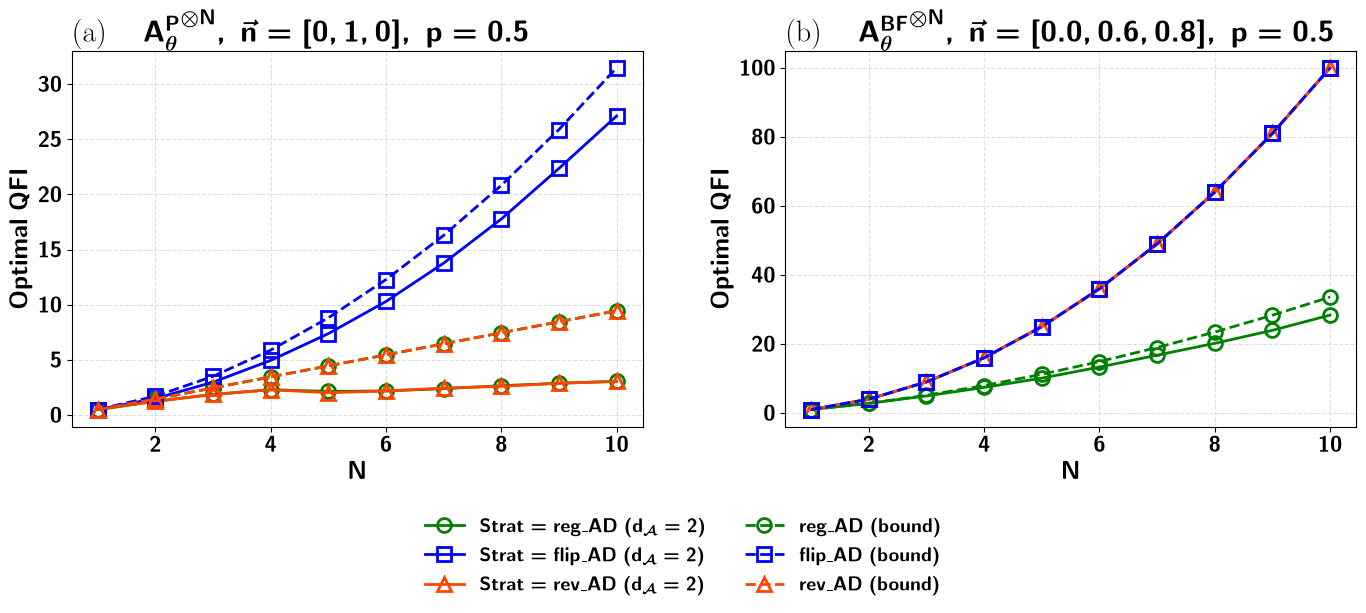}
    \caption{\textbf{ Change of scaling in optimal QFI  with fixed ancillary dimensions $d_{\mathcal{A}} = 2$ and upper bounds (ordinate) with respect to teh number of  usages, \(N\) (abscissa) for various strategies and channels with fixed parameter values.}  For a channel affected by (a) Pauli noise $A^{P}_\theta = \mathcal{U}_\theta \circ A^{P}$ (b) bit-flip noise $A^{BF}_\theta = \mathcal{U}_\theta \circ A^{BF}$. All axes are dimensionless.  }
    \label{fig:finite_pauliBF_N}
\end{figure*}

Similarly, we can demonstrate metrological activation for the transpose supermap or $\textbf{R}$-assisted metrology by considering bit-flip noise $A^{BF}$ with Kraus operators, \begin{equation}
K_1^{BF} = \sqrt{p}I, ~ K_2^{BF} = \sqrt{1-p}\sigma_x,~ 0 \leq p \leq 1,
\end{equation}
and $p = 0.5$. Again, the task is to estimate, $\theta$ in $A^{BF}_\theta = \mathcal{U}_\theta  \circ A^{BF}$ at $\theta = \pi/2$ and $\vec{n} = [0.0,0.6,0.8]$.  The asymptotic upper bounds for various strategies are depicted in Fig.~\ref{fig:asymp_all_N_2}(b). We observe that $\mathcal{J}^{\textbf{flip\_Strat}} = \mathcal{J}^{\textbf{rev\_Strat}} \sim N^2,$ whereas $\mathcal{J}^{\textbf{reg\_Strat}} \sim N$ with $||\beta|| \sim 10^{-10}$. Thus we have a metrological activation. The activation is also present in the finite usage regime. When we perform the ISS tensor network algorithm for Pauli and bit-flip noise metrological tasks with a fixed auxiliary dimension $d_{\mathcal{A}} = 2,$ a distinction in QFI scaling emerges even at finite usage of channels, see Fig.~\ref{fig:finite_pauliBF_N}. Further advantages on bit-flip noise are present in Appendix~\ref{app:bitflip}.

Let us now exhibit that the metrological activation has several implications on the simulability of supermaps.
\begin{proposition}
The quantum time-flip supermap cannot be exactly and deterministically simulated with a quantum comb or causal superposition strategy along with any finite queries of input channels.  
\label{th:flip_unsim}
\end{proposition}
\begin{proof}
    Given $A_\theta^P$ with $\vec n =\hat y$, in the asymptotic region, we have $\mathcal J^\textbf{flip\_Strat}\equiv \mathcal J^\textbf{Strat}({\mathsf T_{\mathsf A_\theta^P}}^{\otimes N})=p_y^2 N^2$ and $\mathcal J^\textbf{reg\_Strat}\equiv\mathcal J^\textbf{Strat}({\mathsf A_\theta^P}^{\otimes N})= N$. 

    Let us assume that $T_{A_\theta^P}$ is deterministically and exactly simulable. This means that there must exist some quantum comb $\mathsf B$ and some finite number $M<N$, $N$ being a multiple $\kappa M$ of $M$ such that ${\mathsf T_{\mathsf A_\theta^P}}=\mathsf B\star {\mathsf A_\theta^P}^{\otimes M}$. Let, $\mathsf B_\text{opt}\in \textbf{Strat}$ be the optimal quantum comb for ${\mathsf T_{\mathsf A_\theta^P}}^{\otimes \frac{N}{M}}$ to get QFI, i.e., 
    \begin{eqnarray}
        \mathcal J^\textbf{Strat}({\mathsf T_{\mathsf A_\theta^P}}^{\otimes \frac NM})=\mathcal J(\mathsf{B}_\text{opt}\star{\mathsf T_{\mathsf A_\theta^P}}^{\otimes \frac NM}).
    \end{eqnarray}
    Therefore, we can write
    \begin{eqnarray}
        \nonumber \mathcal J^\textbf{Strat}({\mathsf A_\theta^P}^{\otimes N})&\geq&\mathcal J(\mathsf B_\text{opt}\star \mathsf B^{\otimes \frac NM}\star{\mathsf A_\theta^P}^{\otimes M \frac NM}),\\\nonumber
        &=&\mathcal J(\mathsf B_\text{opt}\star (\mathsf B\star{\mathsf A_\theta^P}^{\otimes M})^{\otimes \frac NM}),\\\nonumber
        &=& \mathcal J(\mathsf B_\text{opt}\star ({\mathsf T_{\mathsf A_\theta^P}})^{\otimes \frac NM}),\\\nonumber
        &=& \mathcal J^\textbf{Strat}({\mathsf T_{\mathsf A_\theta^P}}^{\otimes \frac NM}).
    \end{eqnarray}
    Hence, the inequality $N> p_y^2\frac{N^2}{M^2} \implies \frac{M}{\kappa}\geq p_y^2$ must hold. However, in the limit $\kappa\to\infty,$ this reduces to $ p_y^2 \leq 0$ which is infeasible since $p_y> 0$. We can prove the theorem for casual superposition \textbf{CS} strategies identically, since they saturate to the same QFI values as adaptive \textbf{AD} strategies in the asymptotic limit. 
\end{proof}

By employing the same line of reasoning as in the proof of Proposition~\ref{th:flip_unsim} but with the bit-flip noise estimation task, if the transpose map is simulable by using a finite number of $M<N$ copies of $A_\theta^{BF},$ the inequality $M/\kappa\geq 1$ must necessarily be satisfied. But this is again violated in the limit $\kappa\to\infty$. Hence, we arrive at the following result:
\begin{proposition}
The transpose supermap, e.g., a supermap which takes as input a channel $ A_\theta$ with Kraus $K_{i, \theta}$ and outputs a channel $ A_\theta^T$ with Kraus $K^T_{i, \theta}$ cannot be exactly and deterministically simulated with a quantum comb or causal superpositions along with any finite queries to input channels.  
\end{proposition}

Interestingly, it has been shown that it is possible to construct a transposition and inversion supermaps for unitaries with just four queries, see Refs. \cite{Yoshida2023Sep, Grinko2024Dec, Odake2024May}. A similar separation in simulability for unitaries and channels is also observed for the quantum switch supermap. While, for unitary inputs, two queries of one of the input unitary allows for a deterministic and exact simulation of a quantum switch \cite{chiribella2013quantum}, for channels, it has been proven that at least an exponential number of queries are required \cite{Kristjansson2024Sep}.


\section{Conclusion}
\label{sec:conclu}

Quantum metrology utilizes quantum resources like coherence and entanglement to achieve measurement precision beyond classical limits. However, in the presence of unavoidable system-environment interactions, the benefit gets diminished, and hence it is essential to search for resources that can remain advantageous even in noisy settings. Our work presents  a rigorous investigation of  metrological protocols employing  quantum operations with indefinite time direction. By focusing on strategies assisted by the quantum time-flip  supermap, we  demonstrated that these protocols fundamentally challenge the well-established hierarchy of known protocols and limitations of quantum metrology.

Using semi-definite programming and asymptotically tight upper bounds,  we found that the time-flip-assisted protocols can achieve a quantum Fisher information (QFI) that is strictly greater than the theoretical maximum attainable by any adaptive, parallel, and causal superposition strategies. Importantly, this advantage is not only present for a finite number of channel usages but also persists in the asymptotic limit, thereby violating the well-established asymptotic equivalence~\cite{Kurdzialek2023Aug}. Hence, 
time-flip-assisted protocols are the only discrete-dimensional metrological methods that demonstrate a genuine asymptotic superiority over parallel schemes  in the limit of a large number of channel queries.


 We found a class of Pauli noise channels for which all conventional metrological strategies, including the most general ones with causal superpositions, are fundamentally limited to the standard scaling. In contrast, time-flip-assisted protocols activate Heisenberg scaling for these same channels, demonstrating a qualitative enhancement in estimation precision. We refer to this phenomenon as \textit{metrological activation}, establishing indefinite time direction as a genuine resource for quantum metrology.  We  used this phenomenon to provide a rigorous proof that the quantum time-flip supermap cannot be simulated by conventional quantum circuits and those with causal superpositions. Moreover, this method of using metrological scaling  offers a new operational tool for certifying the unsimulability of higher-order quantum operations and also proves the unsimulability of the transpose supermap for channels.


Our analysis opens several directions for future research, including a complete characterization of the classes of channels, permitting metrological activation, which could lead to new design principles for noise-resilient quantum sensors. It is also natural to explore more general higher-order transformations of bidirectional processes \cite{Apadula_2026} in noisy metrology.
Our results  establish the quantum time-flip, and more broadly, indefinite time direction as a valuable resource for noise-affected metrology, potentially opening up a new avenue for precise measurements.

\acknowledgements
GA thanks Matt Wilson and Marco Túlio Quintino for fruitful discussions on indefinite time direction and simulability of quantum supermaps. We acknowledge the use of Qmetro++ \cite{Dulian2025Jun} and QuTiP \cite{Johansson2012Aug} for computing optimal QFIs and metrological upper bounds  and \href{https://github.com/titaschanda/QIClib}{QIClib} -- a modern C++ library for general purpose quantum information processing and quantum computing (\url{https://titaschanda.github.io/QIClib}) and cluster computing facility at Harish-Chandra Research Institute. PH acknowledges ``INFOSYS
scholarship for senior students". We acknowledge the support from the project entitled `` Technology Vertical - Quantum Communication'' under the National Quantum Mission of the Department of Science and Technology (DST)  (Sanction Order No. DST/QTC/NQM/QComm/2024/2 (G)).

\bibliography{reference}

\appendix

\section{Derivation of $\mathcal J^{ \textbf{Strat}}(\mathsf C_\theta)$ in MOP Method}
\label{sec:MOP}

Let $\rho_\theta\in\mathcal L(\mathcal H)$ be the quantum state with the parameter $\theta$ encoded in it. The QFI of $\rho_\theta$ with respect to the parameter $\theta$ is given by
\begin{eqnarray}
    \mathcal J(\rho_\theta)=\min_{\ket{\Psi_\theta}:\text{Tr}_A(\ketbra{\Psi_\theta}{\Psi_\theta})=\rho_\theta}4\braket{\dot\Psi_\theta},
    \label{eq:purif}
\end{eqnarray}
where the minimization is done over all possible purifications $\ket{\Psi_\theta}\in\mathcal{H}\otimes\mathcal{A}$ of $\rho_\theta$\cite{Fujiwara2008May,liu2023optimal,Liu2024Dec} and $|\dot\Psi_\theta\rangle=\partial_\theta\ket{\Psi_\theta}$. Here, $\mathcal{H}(\mathcal{A})$ denotes the Hilbert space associated with the system (auxiliary) space, while $\mathcal{L}(X)$ represents the set of density operators defined on the Hilbert space $X$. 

Now, given a completely positive and trace-preserving (CPTP) operation $ A_\theta:\mathcal{L}(\mathcal H)\to\mathcal{L}(\mathcal K)$, estimating $\theta$ with the optimal protocol can be translated into finding the channel QFI 
\begin{eqnarray}
    \mathcal J( \mathsf{A}_\theta)=\max_{{\ket\Phi}_{\mathcal {HR}}}\mathcal{J}[ A_\theta\otimes\mathcal I_{\mathcal R}(\ketbra\Phi)]=\max_{{\ket\Phi}_{\mathcal {HR}}}\mathcal{J}({\rho_\theta}),
\end{eqnarray}
where the pure state $\ket\Phi\in\mathcal H\otimes\mathcal R$ is chosen due to convexity of QFI and $\rho_\theta= A_\theta\otimes\mathcal I_{\mathcal R}(\ketbra\Phi)$. Here, $\mathcal{R}$ denotes an auxiliary Hilbert space on which the channel acts trivially (since $\mathcal I_{\mathcal R}$ is the identity map on $\mathcal R$), while allowing entanglement with the probe system.  Using Eq.~\eqref{eq:purif} we can write 
\begin{eqnarray}
    \mathcal J( \mathsf{A}_\theta) = \max_{{\ket\Phi}_{\mathcal {HR}}}\min_{\ket{\Psi_\theta}_{\mathcal{HRA}}}4\braket{\dot\Psi_\theta}.
\end{eqnarray}

In a pedagogical metrological task, given multiple queries of $ A_\theta$, we need to find the optimal protocol, which may consist of quantum control, error correction, feedback, etc. Mathematically, we need to find the optimal superoperator $P:\mathcal L(\otimes_{i=1}^{N}\mathcal K_i)\to \mathcal L(\otimes_{i=1}^N\mathcal H_i\otimes \mathcal R_N)$ (see Fig.~\ref{fig:mop}) such that its concatenation with $ A_\theta^{\otimes N}$ delivers the state $\rho_\theta^N$ with maximum QFI. To characterize $P$, let us denote the corresponding Choi–Jamiołkowski (CJ) operator as $\mathsf{P}=P\otimes\mathcal I_{\mathcal H_{in}}(\ketbra{\mathbb 1})\in \mathcal L(\mathcal H_1\otimes \mathcal K_1\otimes\mathcal H_2\otimes\mathcal K_2\otimes\ldots\otimes\mathcal K_{N-1}\otimes \mathcal H_N\otimes \mathcal{K}_N \otimes \mathcal R_N)$ where $\mathcal H_{in}\equiv\otimes_{i=1}^{N}\mathcal K_i$ and $\ket{\mathbb 1}=\sum_{j}\ket{j}_{\mathcal{H}_{in}}\ket{j}_{\mathcal{H}_{in}}$ is the unnormalized maximally entangled state in $\mathcal H_{in}\otimes\mathcal{H}_{in}$.

Under this formalism the output state is given by $\rho_\theta^N=\mathsf C_\theta\star\mathsf P$ with $\mathsf C_\theta=[ A_\theta\otimes\mathcal I(\ketbra{\mathbb 1})]^{\otimes N}$ = $\sum_{i=1}^{r^N}| {\mathsf K}^N_{i,\theta} \rangle \langle {\mathsf K}^N_{i,\theta}|$ being the Choi representation of $ A_\theta^{\otimes N}$ and $r$ as the rank of Kraus representation of $ A_\theta$. Here `$\star$' is the link product defined for two arbitrary operator $X\in\mathcal L(\mathcal A_1\otimes\mathcal A_C)$ and $Y\in\mathcal L(\mathcal A_C\otimes\mathcal A_2)$ as $X\star Y=\text{Tr}_{\mathcal A_C}[(X\otimes \mathbb I_{\mathcal A_2})(\mathbb I_{\mathcal A_1}\otimes Y^{T_{\mathcal A_C}})]$ with $T_{\mathcal A_C}$ being the transpose operation with respect to $\mathcal A_C$. Therefore, given a set of strategies $ \textbf{Strat}\equiv\{\mathsf P|\mathsf P\geq 0, \text{rank}(\mathsf P)=1, \text{Tr}_{\mathcal R_N}\mathsf P=\widetilde{\mathsf{P}}\in\widetilde{\textbf{Strat}}\}$, the maximum QFI can be written as \cite{liu2023optimal}
\begin{eqnarray}
    \nonumber \mathcal J^{ \textbf{Strat}}(\mathsf C_\theta)=\max_{\mathsf P\in \textbf{Strat}}\mathcal J(\mathsf C_\theta\star\mathsf P)=4\min_h\max_{\widetilde P\in \widetilde{\textbf{Strat}}}\text{Tr}[\bar{\Omega}(h)\widetilde{\mathsf{P}}],\\
\end{eqnarray}
where ${\Omega}(h)=\sum_i| \dot{\mathsf K}^N_{i,\theta}(h) \rangle \langle \dot{\mathsf K}^N_{i,\theta}(h)|^T$ with the Kraus operators of the single channel satisfying $\dot{\mathsf K}_{i,\theta}(h) = \dot{\mathsf K}_{i,\theta} -i\sum_j h_{ji} \mathsf K_{\theta, j}$ and $h$ is an $r$-dimensional arbitrary hermitian matrix for some fixed Kraus representation corresponding to ${\mathsf K}_{i,\theta}$.

\section{Mathematical Definition of the Strategy Sets}
\label{sec:appendix_strategies}

A generic $N$-step pure strategy is faithfully described by a Choi--Jamiołkowski (CJ) operator $\mathsf{P} \in \mathcal{L}(\bigotimes_{i=1}^N(\mathcal{H}_i \otimes \mathcal{K}_i) \otimes \mathcal{R}_N)$. To precisely express the causality constraints, we follow the left-subscript notation to represent the trace-replace over any subsystem $\mathcal{A}$:
\begin{equation}
    _{\mathcal{A}}\mathsf{C} := \frac{\mathbb{I}_{\mathcal{A}}}{d_{\mathcal{A}}} \otimes \text{Tr}_{\mathcal{A}} \mathsf{C}.
\end{equation}

All valid pure strategy sets $\textbf{Strat} \in \{\textbf{Par}, \textbf{AD}, \textbf{CS}, \textbf{ICO}\}$ share a few base requirements
\begin{equation}
    \mathsf{P} \ge 0, \quad \text{rank}(\mathsf{P}) = 1, \quad \text{Tr}(\mathsf{P}) = \prod_{i=1}^N d_{\mathcal{K}_i},
\end{equation} above which the specific strategies have constraints as follows:

\textbf{Parallel Strategies (Par).} In a parallel strategy, all $N$ processes are queried simultaneously. The inputs cannot depend on any channel outputs. Mathematically $\textbf{Par}$ is defined as the collection of operators $\mathsf{P}$ satisfying the base requirements and
\begin{equation}
    _{\mathcal{R}_N} \mathsf{P} = _{\mathcal{R}_N, \mathcal{K}_1, \mathcal{K}_2, \dots, \mathcal{K}_N} \mathsf{P}.
\end{equation}

\textbf{Adaptive Strategies (AD).}
Adaptive strategies query the $N$ channels in a strict chronological order. Assuming the canonical order $\pi = (1 \to 2 \to \dots \to N)$, causality dictates that operations up to step $k$ cannot depend on future channel outputs. They are defined by the base requirements and the sequential comb constraints
\begin{eqnarray}
    _{\mathcal{R}_N} \mathsf{P} &=& _{\mathcal{R}_N, \mathcal{K}_{N}} \mathsf{P}, \nonumber \\
    _{\mathcal{R}_N, \mathcal{K}_{N},\mathcal{H}_{N} \dots,\mathcal{H}_{i+1}, \mathcal{K}_{i}} \mathsf{P} &=& _{\mathcal{R}_N, \mathcal{K}_{N},\mathcal{H}_{N} \dots,\mathcal{H}_{i+1}} \mathsf{P},\\ \nonumber
    &&\quad \text{for } i = 1, \dots, N - 1.
\end{eqnarray}

\textbf{Causal Superposition Strategies (CS).} This set allows for pure coherent superpositions of different fixed causal orders. Mathematically, it is defined as the collection of operators $\mathsf{P}$ satisfying the base requirements and
\begin{eqnarray}
    \text{Tr}_{\mathcal{R}_N} \mathsf{P} = \sum_{\pi \in S_N} q_\pi \text{Tr}_{\mathcal{R}_N}\mathsf{P}^\pi, \quad \sum_{\pi \in S_N} q_\pi &=& 1, \quad q_\pi \ge 0, \\ \nonumber
    \quad &&\text{with } \mathsf{P}^\pi \in \textbf{AD}^\pi, 
\end{eqnarray}
where $S_N$ is the symmetric group of degree $N$, and $\textbf{AD}^\pi$ denotes the adaptive strategy set executed in the order of permutation $\pi$.

\textbf{General Indefinite Causal Order (ICO).}
For $N=2$, a general indefinite-causal-order strategy set $\textbf{ICO}$ is defined as the collection of operators $\mathsf{P}$ satisfying the base requirements and
\begin{eqnarray}
    _{\mathcal{R}_2, \mathcal{H}_1, \mathcal{K}_1} \mathsf{P} &=& _{\mathcal{R}_2, \mathcal{H}_1, \mathcal{K}_1, \mathcal{K}_2} \mathsf{P}, \nonumber \\
    _{\mathcal{R}_2, \mathcal{H}_2, \mathcal{K}_2} \mathsf{P} &=& _{\mathcal{R}_2, \mathcal{K}_1, \mathcal{H}_2, \mathcal{K}_2} \mathsf{P}, \nonumber \\
    _{\mathcal{R}_2} \mathsf{P} &=& _{\mathcal{R}_2, \mathcal{K}_2} \mathsf{P} + _{\mathcal{R}_2, \mathcal{K}_1} \mathsf{P} - _{\mathcal{R}_2, \mathcal{K}_1, \mathcal{K}_2} \mathsf{P},
\end{eqnarray}
with similar multiparty process matrices defined using projections, see Ref. \cite{Oreshkov2016Sep}. Alternatively, since the only constraint over such a process is that its compositions with channels yields a channel, (In our case a state), we have 
\begin{eqnarray}
\rho_{\mathcal{R}} = \mathsf{P} \star (\otimes_{i =1}^N \mathsf{C}_i), ~\rho_{\mathcal{R}}\geq 0, ~\text{Tr}\rho_{\mathcal{R}} = 1,
\end{eqnarray}
where $\mathsf{C}_i $ are Choi matrices of channels $C_i \in \mathcal{L}(\mathcal{H}_i \otimes \mathcal{K}_i \otimes \mathcal{A}_i)$ composed with the process matrix $\mathsf{P}$ with $\mathcal{A}_i$ being the auxiliaries.

\section{Iterative see-saw (ISS) Method}
\label{sec:ISS}
Although MOP is the most powerful path to calculate optimal QFI, as the number of channel usage or the dimensionality of the system increases, this method becomes inefficient.

Given a parametrized comb $\mathsf{C}_\theta$, for ISS method a pre-QFI function is defined on the strategy $\mathsf{P}$
\begin{eqnarray}
\mathcal{F}(\mathsf{C}_\theta, \mathsf{P},  \mathsf{L}) = 2\Tr(\dot\rho_\theta \mathsf{L}) - \Tr(\rho_\theta \mathsf{L}^2),
\end{eqnarray}
where $\rho_\theta = \mathsf{C}_\theta \star \mathsf{P}$, and $\mathsf{L}$ is a Hermitian operator. Maximizing this quantity over $\mathsf{L}$ gives the QFI for the particular encoded state $\rho_\theta$, where the maximizing Hermitian operator $\mathsf{L}_{opt}$ turns out to be the symmetric logarithmic derivative (SLD).
\begin{eqnarray}
\max_\mathsf{L} \mathcal{F}(\mathsf{C}_\theta, \mathsf{P},  \mathsf{L}) = \mathcal{J}(\mathsf{C}_\theta \star \mathsf{P}).
\end{eqnarray}

Now, we can also maximize this function over the strategies $\mathsf{P}$ to convert optimal metrology into a double maximization problem
\begin{eqnarray}
\mathcal J^{{\textbf{Strat}}}(\mathsf C_\theta) = \max_{\mathsf{P} \in \mathbf{Strat},L}2\Tr(\dot\rho_\theta \mathsf{L}) - \Tr(\rho_\theta \mathsf{L}^2).
\end{eqnarray}
The operators $\mathsf{P}$ and $\mathsf{L}$ can be maximized over one after other iteratively, feeding the fixed maximum $\mathsf{P}~ ~ (\mathsf{L})$ of the previous round to the maximization problem of $\mathsf{L}~~(\mathsf{P})$ in the current round. In the first round we choose a random $\mathsf{P}$ from our strategy set, obtain $\rho_\theta$ and derive $\mathsf{L}$ by solving linear equations $$\dot \rho_\theta = \frac{1}{2}[\rho_\theta \mathsf{L} + \mathsf{L} \rho_\theta].$$
Once, $\mathsf{L}_{opt}$ e.g., SLD operator is received the following SDP can be solved for maximizing over $\mathsf{P}$ for a fixed $\mathsf{L}$
\begin{eqnarray}
\max_\mathsf{P} \mathcal{F}(\mathsf{C}_\theta, \mathsf{P},  \mathsf{L}) &=& \max_\mathsf{P}\text{Tr}(\mathsf{P}\mathsf{M}) \notag\\
\text{where,} \quad \mathsf{M} = 2\mathsf{L} \star \dot{\mathsf{C}}_\theta^T &-& \mathsf{L}^2 \mathsf{C}_\theta^T.
\label{eqn:iss_eqn}
\end{eqnarray}
This procedure is then repeated for any number of rounds until negligible changes of QFI in subsequent rounds are observed. While this technique is advantageous in its own right since, unlike MOP, it allows for optimizations of other Loss functions such as Bayesian quadratic costs thus enabling optimizations for Bayesian metrology as well. For the purpose of our current work the main advantage of the ISS method comes from its ability to be moulded for adaptive metrology in a way that a tensor network approach can be utilized which allows for QFI estimations for high number of channel uses. A quantum comb $P$ can always be written as a series of isometries $\{P_i:\mathcal{K}_{i-1} \otimes \mathcal{A}_{i-1} \to \mathcal{H}_{i} \otimes \mathcal{A}_i\}$, also known as the teeth of quantum combs with ancillary systems $\mathcal{A}_i$ in between and systems $\mathcal{K}_{i-1}, \mathcal{H}_{i}$ which may accept quantum channels or other quantum combs. Then, the Choi for the overall comb $\mathsf{P}$ can be written as a network of these isometries by link product $\mathsf{P} = \mathsf{P}_1 \star \mathsf{P}_2 \star \dots  \star \mathsf{P}_N \star \mathsf{P}_{N+1}$. Once we fix the ancillary dimensions $d_{\mathcal{A}_i} = d_{\mathcal{A}}$ and the systems dimension $d_{\mathcal{K}_i} = d_{\mathcal{H}_i} = d_{\mathcal{H}}$, the optimization over a whole quantum comb can be written as optimization over individual tooth, thus reducing the number of variables to be optimized. While earlier, the number of variables scaled as $d_{\mathcal{H}}^{4N+2} d_{\mathcal{A}}^2$, after breaking into network of isometries the number of variables are $d_{\mathcal{H}}^{2} d_{\mathcal{A}}^2 +Nd_{\mathcal{H}}^{4} d_{\mathcal{A}}^4.$ The maximization becomes $\max_{\mathsf{P}_1, \mathsf{P}_2,\dots,\mathsf{P}_N, \mathsf{P}_{N+1}}\mathcal{F}(\mathsf{C}_\theta, \mathsf{P},  \mathsf{L})$ which demands for a modified ISS approach. Initially $\mathsf{P}_1, \mathsf{P}_2,\dots,\mathsf{P}_N, \mathsf{P}_{N+1},\mathsf{L}$ are initialized randomly, then each $\mathsf{P}_i$ is optimized one by one using Eq.~\eqref{eqn:iss_eqn}, keeping all other teeth as well as operator $\mathsf{L}$ to be fixed. Lastly, $\mathsf{L}$ is optimized by solving the linear system of equations. This procedure is repeated until desired accuracy is achieved. In general, for low ancillary dimension $d_{\mathcal{A}}$ this procedure allows for a significantly higher number of channel usage to be optimized.

\begin{figure*}[!t]
    \centering
\includegraphics[width=1\textwidth]{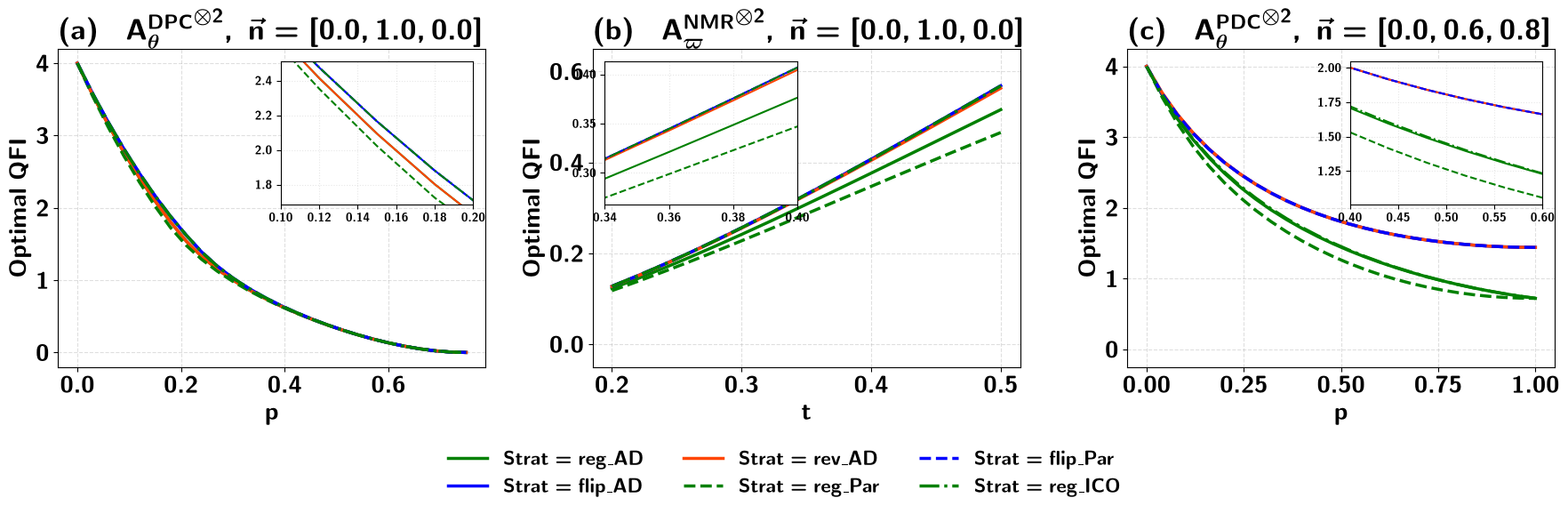}
\caption{\textbf{ Optimal QFI  (ordinate) vs noise parameters (abscissa) for various strategies and channels with finite usage $N = 2.$} (a) For channel affected by depolarizing noise $A^{DPC}_\theta = \mathcal{U}_\theta \circ A^{DPC}$. (b) For channel affected by NMR noise $A^{NMR}_\varpi = \mathcal{U}_\varpi \circ A^{NMR}$. (c) For channel affected by phase damping noise $A^{PDC}_\theta = \mathcal{U}_\theta \circ A^{PDC}$. All axes are dimensionless.} 
    \label{fig:finite_all_param_n2}
\end{figure*}

\section{Optimal QFI for $N=2$}
\label{app:optimalqfin2}
In this section we discuss optimal QFI evaluated for $N = 2$ usage of channels across the three prototypical noise models: depolarizing channel (DPC), nuclear magnetic resonance (NMR) noise, and phase damping channel (PDC) exactly like the $N=3$ case where noise $A^{NOISE}$ is applied to the probe before the parametrized unitary $\mathcal{U}_\theta = e^{i \vec{n} \cdot \vec{\sigma} \theta/2}(*)e^{-i \vec{n} \cdot \vec{\sigma} \theta/2}$. For $N=2$, the exact Maximum Over Purification (MOP) algorithm is highly efficient and computationally tractable. Consequently, unlike the $N \ge 3$ regimes where the iterative see-saw (ISS) tensor network method was employed with a constrained auxiliary dimension ($d_{\mathcal{A}}$), the $N=2$ optimal QFIs presented here are unconditional and exact, evaluated over the full strategy space. 

As illustrated in Fig.~\ref{fig:finite_all_param_n2}, the fundamental hierarchies between the strategy sets begin to emerge even at the lowest multi-query level, albeit with narrower separations than those observed at higher $N$.

\noindent\textit{Depolarizing Noise.} For the estimation of $\theta=\pi/2$ under depolarizing noise (with $\hat{n} = \hat{y}$), we evaluate the optimal QFI as a function of the noise strength $p$ [Fig.~\ref{fig:finite_all_param_n2}(a)]. At $N=2$, the performance gap between the strategies is marginal but strict. Notably, \textbf{reg\_Par} performs almost as well as the other strategies at low noise levels, but slight improvements are provided by \textbf{reg\_AD} and \textbf{rev\_AD}, which are equivalent ($\textbf{reg\_AD} \equiv \textbf{rev\_AD}$) due to the symmetry of the depolarizing channel. The indefinite causal order strategy \textbf{reg\_ICO} and the time-flip-assisted strategies \textbf{flip\_AD} and \textbf{flip\_PAR} offer marginal improvements over the definite causal order schemes. Although the genuine metrological advantage of time-flip (GMATF) is minuscule at $N=2$, its presence confirms that the operations with indefinite time direction act as a metrological resource even for minimal channel usages, foreshadowing the more substantial advantage seen at higher $N$.

\noindent\textit{NMR Noise.} In the case of the NMR noise model, we estimate the frequency $\varpi = 10$ kHz (with $\hat{n} = \hat{y}$ and $t \in [0, 1.2]$s) [Fig.~\ref{fig:finite_all_param_n2}(b)]. The hierarchy of the strategies is muted in comparison to the $N=3$ case because of the small strategy space. We observe $\textbf{flip\_AD} \equiv \textbf{flip\_Par}  \equiv \textbf{reg\_ICO} >  \textbf{rev\_AD} > \textbf{reg\_AD} > \textbf{reg\_Par}$. Crucially, the GMATF is available even at $N=2$, manifesting as a slight gap between the QFI of \textbf{flip\_Strat} and the input-output inverted \textbf{rev\_Strat}. The gap monotonically increases with the evolution time $t$, validating the benefit of TF-assisted metrology.

\noindent\textit{Phase Damping Noise.} For the dephasing noise scenario, estimating $\theta = \pi/2$ along the axis $\hat{n} = [0, 0.6, 0.8]$, we observe that the QFI curves for \textbf{flip\_AD} and \textbf{rev\_AD} identically overlap ($\textbf{flip\_AD} \equiv \textbf{rev\_AD}$) [Fig.~\ref{fig:finite_all_param_n2}(c)]. Both of these strategies significantly outperform their regular counterparts (\textbf{reg\_AD} and \textbf{reg\_Par}) at high noise levels ($p \to 1$). Because the time-flip advantage is perfectly matched by the time-reversal protocol, the performance enhancement is entirely attributable to the input-output inversion rather than a true superposition of time directions. Specifically, the transpose operation effectively applies the dephasing noise after the signal unitary, enabling an unperturbed parameter encoding that boosts the QFI.

\section{Other advantages in Bit-flip Noise}
\label{app:bitflip}
\begin{figure}[t]
  \centering
  \includegraphics[width=0.98\columnwidth]{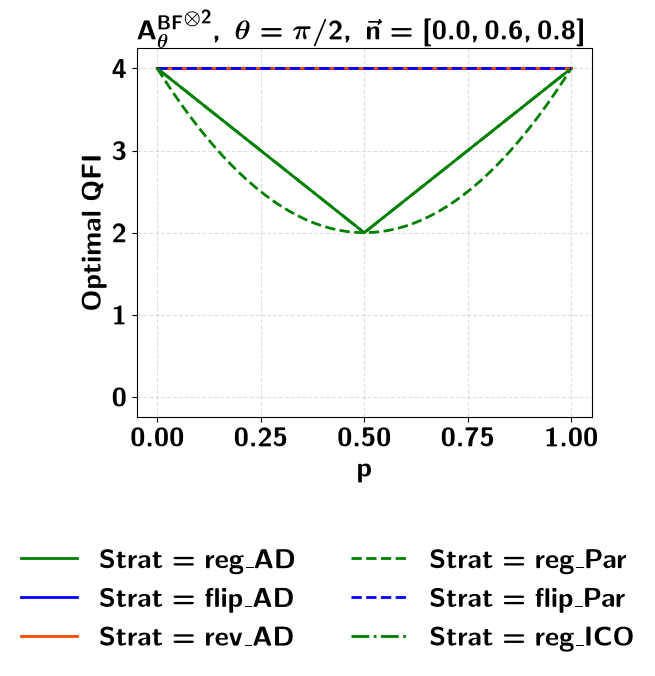}
    \caption{\textbf{ Optimal QFIs  (ordinate) for various strategies under bit-flip noise with strength $p$ (abscissa) at $N=2$.} We observe substantial improvements with both input-output inversion and quantum time-flip. The inversion (transpose) effectively ensures that the unitary acts prior to the bit-flip noise, allowing the protocol to attain perfect $N^2$ scaling behavior across all noise levels. All axes are dimensionless.}
    \label{fig:finite_BF_param}
\end{figure}
Here we present a detailed analysis of the protocol's performance under bit-flip noise. Akin to the phase damping scenario, we observe that the substantial metrological advantages conferred by the quantum time-flip under bit-flip noise can be entirely attributed to the input-output inversion operation.

The estimation task focuses on the phase parameter $\theta$ (set to $\theta = \pi/2$) encoded via the unitary $U_\theta = e^{i\theta \vec{n}\cdot\vec{\sigma}/2}$ along the specific axis $\vec{n} = [0.0, 0.6, 0.8]$. The environment induces a bit-flip noise acting immediately before the unitary, characterized by the Kraus operators $K^{BF}_1 = \sqrt{p}I$ and $K^{BF}_2 = \sqrt{1-p} \sigma_x$.

The behavior of the optimal QFI at $N=2$, calculated via the exact MOP method, is illustrated in Fig.~\ref{fig:finite_BF_param}. We observe a dramatic improvement in precision when employing both the \textbf{R}-assisted (\textbf{rev\_Strat}) and \textbf{TF}-assisted (\textbf{flip\_Strat}) protocols compared to the regular strategies. The physical mechanism behind this advantage relies on the action of the transpose map: by inverting the input-output direction, the noise effectively acts on the probe \textit{after} the unitary encoding has occurred. This chronological inversion allows the metrological protocol to bypass the decoherence during the critical encoding phase, enabling an uncorrupted accumulation of the relative phase and leading to a QFI scaling of $\sim N^2$ regardless of the noise strength $p$.

Furthermore, when optimizing for quantum combs at higher channel usages $N$ using the ISS tensor network algorithm with fixed ancillary dimensions ($d_{\mathcal{A}} = 2$), the behavior observed mirrors that of the dephasing noise, see Fig. \ref{fig:finite_pauliBF_N}. Both time-flip and input-output inversion rigorously outperform the upper bounds of the optimal regular strategies with $\mathcal{J}^{\textbf{flip\_Strat}} \equiv \mathcal{J}^{\textbf{rev\_Strat}}$.

\begin{figure}[t]
  \centering
  \includegraphics[width=0.98\columnwidth]{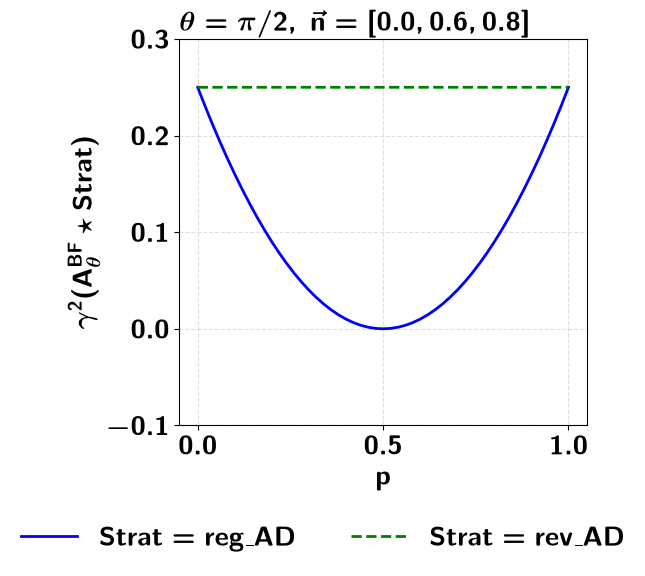}
    \caption{\textbf{${\gamma^2}$ coefficient (ordinate) for R-assisted metrology on channels affected by bit-flip noise with noise parameter $p$ (abscissa).} At $p = 0.5$, the \textbf{reg\_AD} strategy yields a coefficient value of zero, confining it to the standard scaling (SS) regime. On the other hand, the \textbf{rev\_AD} strategy attains a coefficient value of 0.25, pushing the protocol into the Heisenberg scaling (HS) regime. The unsimulability of the transpose supermap follows directly from this activation. All axes are dimensionless.}
    \label{fig:gamma_BF_param}
\end{figure}

To formalize this advantage in the asymptotic regime, we analyze the coefficient $\gamma^2$, which dictates the presence of Heisenberg scaling (HS). In Fig.~\ref{fig:gamma_BF_param}, we plot the $\gamma^2$ for input-output inversion adaptive protocol and the regular adaptive protocol. At the maximum entropy noise level $p = 0.5$, we find that $\gamma^2_{\textbf{reg}} = 0$, meaning the regular adaptive protocol (\textbf{reg\_AD}) is strictly confined to the standard scaling (SS) regime ($\mathcal{J} \propto N$). In stark contrast, the input-output inverted protocol yields $\gamma^2_{\textbf{rev}} = 0.25$, activating the Heisenberg limit ($\mathcal{J} \propto N^2$).

This instance of metrological activation, where a strict change in the fundamental scaling of precision is achieved exclusively via the input-output inversion has implications for its simulability. Following the logic deployed in Proposition \ref{th:flip_unsim}, the fact that \textbf{rev\_Strat} achieves HS while \textbf{reg\_Strat} is restricted to SS provides an operational proof that the transpose supermap cannot be deterministically or exactly simulated by any standard quantum comb or causal superposition utilizing a finite number of queries to the original bit-flip channel.

\end{document}